\documentclass[aip,chaos,reprint,10pt,superscriptaddress,showpacs]{revtex4-1}
\usepackage{amsfonts} 
\usepackage{amsmath}
\usepackage{amssymb}
\usepackage{graphicx}
\usepackage{subfigure}
\usepackage{color}
\usepackage{enumerate}
\usepackage[]{natbib}
\usepackage{sidecap}
\usepackage{soul}
\usepackage{cancel}
\usepackage{ulem}
\usepackage{dcolumn}%
\usepackage{bm}%

\newcommand*\xbar[1]{%
  \hbox{%
    \vbox{%
      \hrule height 0.5pt 
      \kern0.5ex
      \hbox{%
        \kern-0.1em
        \ensuremath{#1}%
        \kern-0.1em
      }%
    }%
  }%
} 

\begin{document}

\title[]{Chaos and complexity in the dynamics of nonlinear  Alfv{\'e}n waves in a magnetoplasma}
\author{Subhrajit Roy}
\email{suvo.math88@gmail.com}
\affiliation{Department of Mathematics, Siksha Bhavana, Visva-Bharati University, Santiniketan-731 235, India}
\author{Animesh Roy}
\email{aroyiitd@gmail.com}
\affiliation{Department of Mathematics, Siksha Bhavana, Visva-Bharati University, Santiniketan-731 235, West Bengal, India}
\author{Amar P. Misra}
\email{apmisra@visva-bharati.ac.in}
\affiliation{Department of Mathematics, Siksha Bhavana, Visva-Bharati University, Santiniketan-731 235, West Bengal, India} 
\date{{\today}}

\begin{abstract}
The nonlinear dynamics of circularly polarized dispersive Alfv{\'e}n wave  (AW) envelopes coupled to the driven ion sound waves  of plasma slow response is studied in a uniform magnetoplasma. By restricting the wave dynamics to a few number of harmonic modes, a low-dimensional dynamical model is proposed to describe the nonlinear wave-wave interactions.  It is found that two subintervals of the wave   number of modulation $k$ of AW envelope exists, namely  $(3/4)k_c<k<k_c$ and $0<k<(3/4)k_c$, where $k_c$ is the critical value of $k$ below which the modulational instability (MI)  occurs.  In the former, where the MI growth rate is low,    the periodic and/or quasi-periodic states are shown to occur, whereas   the latter, where the MI growth is high,  brings about the chaotic states. The existence of these states is established by the analyses of    Lyapunov exponent spectra together with the bifurcation diagram  and phase-space portraits  of dynamical variables.  Furthermore, the   complexities of chaotic phase  spaces in the   nonlinear motion are measured   by  the estimations of  the correlation dimension (CD) as well as the approximate entropy (ApEn), and    compared with those for the known H{\'e}non map and the   Lorenz  system in which  a good qualitative agreement is noted.   The   chaotic motion thus predicted in a low-dimensional model  can be a prerequisite for the onset of Alfv{\'e}nic wave turbulence to be observed in a higher dimensional model that and is relevant  in the Earth's ionosphere and magnetosphere. 
 
\end{abstract}
\maketitle

\textbf{The generation of envelope solitons in the nonlinear interactions  of high-frequency wave electric field and low-frequency plasma density perturbations has been recognized as one of the most important features in the context of plasma heating,  transport of plasma particles, as well as  wave turbulence in modern physics.   One particular class of such   solitons is the   Alfv{\'e}n  solitons  that are circularly polarized high-frequency dispersive waves trapped by the plasma density troughs of low-frequency perturbations. This work proposes a new low-dimensional dynamical model to govern  the nonlinear interactions of these  dispersive Alfv{\'e}n waves with low-frequency plasma density fluctuations and shows  how the nonlinear dynamics can transit from periodic to chaotic states.  The complexity of such chaotic states are also measured by means of the correlation dimension  and the approximate entropy, and  compared with those for the known H{\'e}non map and the   Lorenz  system. The existence of chaos in the evolution of Alfv{\'e}nic wave envelopes can be a good indication for the onset of Alfv{\'e}nic wave turbulence that is relevant in the Earth's ionosphere and magnetosphere.  }
\section{Introduction}\label{sec-intro} 
Alfv{\'e}n waves are typical magnetohydrodynamic (MHD) waves that travel along the magnetic field lines and can be excited in any electrically conducting fluid permeated by a magnetic field.    Such waves can be dispersive in warm electron-ion magnetoplasmas due to the effects of finite ion Larmor radius and the electron pressure gradient force. However, in cold plasmas, they may become dispersive due to finite values of the wave frequency (in comparison with the ion-cyclotron frequency) and the electron inertial force \cite{jana2017}.   Since the theoretical description of their existence by  Alfv{\'e}n in $1942$ \cite{alfven1942} and experimental verification by Lundquist in $1949$ \cite{lundquist1949},   the  Alfv{\'e}n waves (especially with large amplitude) have been   known to play significant roles in  transporting energy and momentum in many geophysical and astrophysical MHD flows including the solar corona and the solar wind. They have also been observed in Earth's magnetosphere \cite{voigt2002}, in interplanetary plasmas \cite{tsurutani1999}, and in solar photosphere \cite{nakariakov1999}, and proposed as the origin of geomagnetic jerks \cite{bloxham2002}. Furthermore, the dispersive Alfv{\'e}n waves (DAWs) can have a wide range of applications in laboratory and space   plasmas  \cite{shukla2004,gekelman1999}. 
 \par
Large amplitude Alfv{\'e}n waves interacting with plasmas can give rise to different nonlinear effects including the parametric  decay of three-wave interactions \cite{shi2017}, stimulated Raman and Brillouin scattering \cite{jain1986}, modulational instability (MI) of wave envelopes \cite{wang2021}, plasma background density modification due to the  Alfv{\'e}n wave ponderomotive force, the  Alfv{\'e}n solitons \cite{einar1986}, as well as the formation of Alfv{\'e}n vortices  and related phenomena \cite{roberts2016,chmyrev1988}   that have been observed in the Earth's ionosphere and magnetosphere. 
For some other important nonlinear effects involving Alfv{\'e}n waves in plasmas, readers are referred to the review work of Shukla and Stenflo \cite{shukla1995}.  
 Furthermore, the formation of envelope solitons associated with  the modulational instability due to the nonlinear interaction of high-frequency wave electric field and low-frequency ion density perturbations has been known to be one of the most important features in the context of chaos and wave turbulence  in plasmas \cite{misra2009,banerjee2010}. When the electric field intensity is so high that the wave number of modulation  exceeds its threshold value,  the envelopes are essentially trapped by the density cavities of plasma slow response and the interactions result into chaos. As this chaotic process develops  in a low-dimensional dynamical system, the rate of transfer (or redistribution) of energy from lower to higher harmonic modes (from large to small spatial length scales) becomes faster, leading to strong wave turbulence. Such  scenarios  have been reported in different contexts by means of Zakharov-like equations in plasmas \cite{zakharov1975,banerjee2010,misra2009,misra2011}.   
\par 
 The nonlinear coupling of circularly polarized dispersive Alfv{\'e}n waves and ion density perturbations associated with plasma slow motion has been studied by Shukla \textit{et al.} \cite{shukla2004} in a uniform magnetoplasma. They proposed a set of coupled nonlinear equations for the wave electric field and the plasma density perturbation which admits a localized DAW envelope accompanied  by a plasma density depression. However, the  theory of nonlinear wave-wave interactions associated with the DAWs  has not been studied yet.  The purpose of the present work is to reconsider this model equations and to study the dynamical features of nonlinear three-wave  interactions numerically in a low-dimensional dynamical model. We show that the transition from order to chaos is  indeed  possible when the wave number of modulation is within the domain of the excitation of three wave modes. The existence of periodic, quasiperiodic and chaotic states is confirmed by inspecting the    Lyapunov exponent spectra, the bifurcation diagram, and phase-space portraits  of dynamical variables.  The   complexities of chaotic phase spaces are also examined  by  the estimations of   correlation dimension (CD) and the approximate entropy (ApEn), and  the obtained results are  compared with those for the H{\'e}non map and the Lorenz system.  A good qualitative agreements of the results are noticed.
\par 
The manuscript is organized as follows:  In Sec. \ref{sec-model}, the modulational instability of AW envelopes is studied and the construction of a low-dimensional dynamical model from a higher dimensional system  is shown. The basic dynamical properties of the low-dimensional system is studied and the existence of periodic, quasi-periodic or chaotic states  is shown in Sec. \ref{sec-dynm-props}.   In Sec. \ref{sec-complexity}.  the complexities of chaotic phase spaces are measured and compared with those for the Lorenz system and H{\'e}non map. Finally, the results are concluded in Sec. \ref{sec-conclu}.
\section{Low-dimensional model} \label{sec-model} 
The nonlinear interactions of circularly polarized  dispersive Alfv{\'e}n wave envelopes propagating along the constant magnetic field $\mathbf{B}_0=B_0\hat{z}$  and the slowly varying electron/ion density fluctuations  that are driven by the Alfv{\'e}n wave ponderomotive force   can be described by the following   set of coupled equations \cite{shukla2004,shukla1986}.
\begin{equation}
\Bigg(\frac{\partial}{\partial t}+V_{A}\frac{\partial}{\partial z}\Bigg)E_\bot-\frac{V_{A}^2}{2n_0}\frac{\partial}{\partial z}(n_1 E_\bot)\pm i\frac{V_{A}^2}{2\omega_{ci}}\frac{\partial^2}{\partial z^2} E_\bot=0,									\label{eq1}
\end{equation}
\begin{equation}
\Bigg(\frac{\partial^2}{\partial t^2}-C_s^2 \frac{\partial^2}{\partial z^2}\Bigg)n_1=-\frac{n_0 V_A^2}{B_0^2} \frac{\partial^2}{\partial z^2} |E_\bot^2|,  \label{eq2}		
\end{equation}
where $E_\bot$ is the perpendicular (to $\hat{z}$) component of the wave electric field, $n_1$ is the plasma number density perturbation (with $n_0$ denoting the equilibrium value),  $V_A=B_0/\sqrt{4\pi n_0m_i}$ is the Alfv{\'e}n velocity,   and $\omega_{ci}=eB_0/cm_i$ is the ion  cyclotron frequency with $e$ denoting the elementary charge, $c$   the speed of light in vacuum, and $m_i$ the ion mass.   Also, $C_s=\sqrt{T_e/m_i}$ is the ion-sound speed with $T_e$ denoting the electron thermal energy. For the description of the linear theory of circularly polarized dispersive Alfv{\'e}n waves and the derivations  of the nonlinear coupled equations \eqref{eq1} and \eqref{eq2},  readers are referred to the work of Shukla \textit{et al.} \cite{shukla2004} 
\par 
By defining the dimensionless quantities according to   
  $t\rightarrow t\omega_{ci},~  z\rightarrow z \omega_{ci}/C_s,~  n\rightarrow n_1/n_0$, and $E\rightarrow cE_\bot /C_sB_0$, Eqs.  \eqref{eq1} and \eqref{eq2} can be reproduced as
  \begin{equation}
\frac{\partial^2 n}{\partial t^2}-\frac{\partial^2 n}{\partial z^2}= -\alpha^2 \frac{\partial^2}{\partial z^2} |E|^2,											\label{eq3}
\end{equation}
\begin{equation}
\frac{\partial E}{\partial t}+\beta \frac{\partial E}{\partial z}-\frac{\beta
}{2} \frac{\partial}{\partial z}(nE)+i \gamma \frac{\partial ^2 E}{\partial z^2}=0,	\label{eq4}
\end{equation}
where $\alpha=V_A /c,~ \beta=V_{A}/C_s$, and $\gamma=\pm\beta^2/2$. Here, the $\pm$ sign in $\gamma$ corresponds to the right- and left-circularly polarized AWs.
\par
Looking for the modulation of the AW amplitude and thereby making  the ansatz,
\begin{equation} \label{eq-ansatz}
\begin{split}
E(z,t)= {E}_1(z,t)\exp[i\theta(z,t)],\\
n(z,t)=\tilde{n}(z,t)\exp\left(ikz-i\omega t\right)+\mathrm{c.c.},\\
{E}_1(z,t)=E_0+\tilde{E}\exp\left(ikz-i\omega t\right)+\mathrm{c.c.},\\
\theta(z,t)=\theta_0+\tilde{\theta}\exp\left(ikz-i\omega t\right)+\mathrm{c.c.}, 
\end{split}
\end{equation}
where $E_1$ and $\theta$ are slowly varying functions of $z$  and $t$, and $\tilde{E}\ll E_0$, $\tilde{\theta}\ll \theta_0$, we obtain from Eqs. \eqref{eq3} and \eqref{eq4} the  following linear dispersion relation for the modulated DAW envelope (For details, see Appendix \ref{appendix-A}).
\begin{equation}
\left(\omega^2-k^2\right)\left[\left(\omega-\beta k\right)^2-\gamma^2k^4\right]+\alpha^2\beta\gamma k^5|E_0|^2=0.
\label{eq-disp}
\end{equation}
For the modulational instability, we assume $\omega\approx \beta k+i\Gamma$ with $\beta k\gg\Gamma$. The instability growth rate is then obtained as 
\begin{equation}
\Gamma=\sqrt{\gamma k^{3}\left(\frac{\beta\alpha^2|E_0|^2}{\beta^2-1}-\gamma k\right)}.
\end{equation} 
Thus, the modulational instability sets in for $0<k<k_c$, where $k_c\equiv  {2\alpha^2|E_0|^2}/{\beta|\left(\beta^2-1\right)|}$ is the critical wave number with $\beta>1$, and the maximum growth rate is attained at $k=(3/4)k_c$. From the expression of $\Gamma$ we find that the growth rate increases in the interval $0\lesssim k\lesssim(3/4)k_c$, reaches maximum at $k=(3/4)k_c$, and then decreases with $k$ with a cutoff at $k=k_c$. However, if the electric field intensity is so high that the MI threshold exceeds the decay instability threshold, the DAWs may be trapped by the ion-sound density perturbations. In this case, the interaction between the circularly polarized DAWs and the ion-sound waves may result into a turbulence  in which the transfer or redistribution of wave energy among different modes can take place \cite{misra2010,misra2009,misra2011}.
On the other hand,  in the adiabatic limit, i.e., the quasi-stationary response of density fluctuations, the second order time derivative in Eq. \eqref{eq3} can be disregarded. The resulting equation is then the derivative NLS (DNLS) equation, given by,
\begin{equation}
\left(\frac{\partial  }{\partial t}+\beta \frac{\partial  }{\partial z}\right)E-\frac{1}{2}\beta \alpha^2
  \frac{\partial}{\partial z}\left(|E|^2E\right)+i \gamma \frac{\partial ^2 E}{\partial z^2}=0.	\label{eq-dnls}
\end{equation} 
Equation \eqref{eq-dnls} is clearly integrable \cite{kawata1968,kaup1978} and hence nonchaotic. So, it can have a localized solution for the wave electric field   envelopes.  
\par 
Equations \eqref{eq3} and \eqref{eq4} are, in general, multidimensional, and can describe the evolution of an infinite number of wave modes. However, a few number modes may be assumed to  participate actively in the nonlinear wave-wave interactions. Such cases are not only common in the Alfv{\'e}nic wave turbulence, but also occur in the parametric instabilities of high- and low-frequency wave interactions close to the instability threshold. In this situation, a low-dimensional model with a few truncated modes is well applicable to study the basic features of the full wave dynamics of Eqs. \eqref{eq3} and \eqref{eq4}. Here, one must note that the specific details of the low-dimensional model strongly depends on the range of the wave number of modulation $k$.  So, considering the nonlinear dynamics among a few number of wave modes, we expand the electric field envelope $E(z,t)$ and the density perturbation $n(z,t)$    as 
\begin{equation}
\begin{split}
E(z,t)=&\sum_{m=-M/2}^{+M/2}E_m(t)e^{imkz}=\sum_{m=-M/2}^{+M/2}\rho_m(t)e^{\theta_m(t)}e^{imkz},\\
=&E_0(t)+E_{-1}(t)  e^{-ik z}+E_{1}(t)  e^{ik z},		\label{eq5}
\end{split}
\end{equation} 
\begin{equation}
\begin{split}
n(z,t)=&\sum_{m=-M/2}^{+M/2}n_m(t)e^{imkz},\\
=&n_0(t)+n_{1}(t)  e^{ik z}+n_{-1}(t)  e^{-ik z},		\label{eq6}
\end{split}
\end{equation}
where $M=[k^{-1}]$ denotes the number of modes to be selected in the interactions,   $E_{-m}=E_{m}$, and $n_{-m}=n_{m}$. For three-wave interactions, we choose $M=2$ and following the same approach as in Refs. \cite{misra2010,misra2010a}, we obtain from Eqs. \eqref{eq3} and \eqref{eq4} the following set of reduced equations. 
 \begin{equation}
\ddot{n}_1-k^2 n_1=\alpha^2 k^2 n_0\sin\psi\cos\phi,						\label{e7}
\end{equation} 	 
\begin{equation}
\dot{\psi}= \beta k n_1 \sin\phi,													\label{e8}
\end{equation}
\begin{equation}
\dot{\phi}= k(\beta -  \gamma k)-\frac{1}{2} \beta k n_1 \tan\frac{\psi}{2} \cos\phi,   \label{e9}
\end{equation}
where the dot denotes differentiation with respect to $t$, $\phi=\theta_0-\theta_1$, {$\psi=2w$}, and  $n_0=|E_{-1}|^2+|E_{1}|^2+|E_0|^2$ is the conserved plasmon number. The detailed derivations of Eqs. \eqref{e7}-\eqref{e9} are given in Appendix \ref{appendix-B}.  The system of Eqs. \eqref{e7} to \eqref{e9} can be recast as an autonomous system: 
\begin{equation}
\begin{split}
&\dot{x}_1= \beta_0 x_2 \sin{x_4},	\\
&\dot{x}_2=x_3,\\
&\dot{x}_3=-k^2 x_2+\alpha_0^2\sin{x_1}\cos{x_4},	\\					 			 
&\dot{x}_4= \gamma_0-\frac{1}{2} \beta_0 x_2 \tan\frac{x_1}{2} \cos{x_4},   \label{eq-system}
\end{split}
\end{equation}
where $\beta_0=\beta k$, $\alpha_0=\alpha k \sqrt{n_0}$, and $\gamma_0=k(\beta -\gamma k)$ of which the key parameters are $\alpha$, $\beta$ and $k$. Also, for the sake of convenience,  we have redefined the variables as $\psi=x_1,~n_1=x_2,~\dot{n}_1=x_3$, and $\phi=x_4$.
\section{Dynamical Properties} \label{sec-dynm-props}  In this section, we numerically study the   linear stability analysis of Eq. \eqref{eq-system} and look for different parameter regimes for the existence of periodic, quasiperiodic and chaotic states  on the basis of lyapunov exponent spectra, bifurcation diagram, and phase-space portraits. 
\subsection{Equilibrium points and eigenvalues} \label{sec-eigen}
 As a starting point,  we  calculate the equilibrium points  by equating the right-hand sides of Eq. \eqref{eq-system} to zero and finding solutions for $x_1,~x_2,~x_3,~x_4$  as   $\left(x_{10},~x_{20},~0,~n\pi\right)$, where $x_{10}=4n\pi\pm2\sin^{-1}\left(\pm\sqrt{{\gamma_0k^2}/{\alpha_0^2 \beta_0}}\right)$ and $x_{20}=\pm(-1)^n\left({2k}/{\beta_0}\right)\sqrt{\left(\beta-\gamma k\right)\left(\alpha^2n_0\beta-\beta+\gamma k\right)}$ with  $n$ being zero or an integer.  Thus, there are primarily  four types of  equilibrium points, namely (For details, see Appendix \ref{appendix-C})  $P_1\equiv\left(x^{+}_{10},~x^{+}_{20},~0,~n\pi\right)$, $P_2\equiv\left(x^-_{10},~x^-_{20},~0,~n\pi\right)$,  $P_3\equiv\left(x^+_{10},~x^-_{20},~0,~n\pi\right)$, and $P_4\equiv\left(x^-_{10},~x^+_{20},~0,~n\pi\right)$, where $x^{\pm}_{10}~(x^{\pm}_{20})$ are the values   corresponding to the $\pm$ signs in  $x_{10}~(x_{20})$ [Here, we are not considering any  sign convention in $\gamma$ applicable to right- or left-circularly polarized AWs]. We note that  $(0,0,0,0)$ is not  an equilibrium point since  for  $(0,0,0,0)$ to be an equilibrium point, one must have $k\sim2/\beta$ which may not satisfy the restriction $k<k_c$ for some typical parameter regimes with $\alpha\ll1$, $E_0>1$, and $\beta>1$.  Furthermore, for real values of $x_{10}$ and $x_{20}$, one must have  $\alpha^2n_0\lesssim1$ and $(2/\beta)\left(1-\alpha^2n_0\right)\lesssim k\lesssim (2/\beta)$.      Next, applying the   the transformation around the equilibrium point, i.e., $x_1^\prime=x_1-x_{10}$, $x_2^\prime=x_2-x_{20}$, $x_3^\prime=x_3-x_{30}$ and $x_4^\prime=x_4-x_{40}$, we  obtain a linearized system of   the form: ${dX^\prime}/{dt}=JX^\prime$ where $J$ is the Jacobian matrix and $X^\prime=(x_1^\prime,~ x_2^\prime,~x_3^\prime,~x_4^\prime)$.  The eigenvalues $(\lambda)$ corresponding to each of these equilibrium points can   be obtained  from  the relation $JX^\prime=\lambda X^\prime$ and then the stability of the system \eqref{eq-system} can   be studied by the nature of these eigenvalues. 
The Jacobian matrix $J$ is given by
\begin{widetext}
\begin{equation}
J = \begin{bmatrix}
0 &\beta_0 \sin{x_4} &0 &\beta_0x_2 \cos{x_4} \\
0 &0 &1 &0  \\
\alpha_0^2\cos{x_1}\cos{x_4}&-k^2  &0 &-\alpha_0^2 \sin{x_1} \sin{x_4}\\
 -\frac{1}{4}\beta_0x_2\sec^2{(\frac{x_1}{2})} \cos{x}_4 &-\frac{1}{2}\beta_0\tan{\left(\frac{x_1}{2}\right)} \cos{x_4}& 0 &\frac{1}{2}\beta_0 x_2 \tan\left(\frac{x_1}{2}\right) \sin{x_4}
\end{bmatrix}, \label{eq-J}
\end{equation}
\end{widetext}
which at the equilibrium point  $\left(x_{10},~x_{20},~0,~n\pi\right)$ reduces to
  \begin{widetext}
  \begin{equation}
J = \begin{bmatrix}
0 &0 &0 &(-1)^n \beta_0x_{20} \\
0 &0 &1 &0  \\
(-1)^n\alpha_0^2\cos{x_{10}}&-k^2  &0 &0\\
 (-1)^{n+1}\frac{1}{4}\beta_0x_{20}\sec^2{(\frac{x_{10}}{2})} &(-1)^{n+1}\frac{1}{2}\beta_0\tan{(\frac{x_{10}}{2})}& 0 &0
\end{bmatrix}, \label{eq-Jac}
\end{equation}
where $n$ is either zero or an integer.
\end{widetext}
The  characteristic equation for the matrix $J$ [Eq. \eqref{eq-Jac}] is 
\begin{equation} 
\lambda^4+\left[k^2+\frac{1}{4}\beta_0^2x_{20}^2\sec^2\left(\frac{x_{10}}{2}\right)\right]\lambda^2+\delta=0,\label{eq-char}
\end{equation}
where 
\begin{equation}
\begin{split}
\delta=&\beta_0x_{20}\left[\frac{1}{2}(-1)^n\alpha_0^2\beta_0\cos\left(x_{10}\right)\tan\left(\frac{x_{10}}{2}\right)\right.\\
&\left.+\frac{k^2}{4}\beta_0x_{20}\sec^2\left(\frac{x_{10}}{2}\right)\right].
\end{split}
\end{equation}
 \par
\begin{figure}
            \includegraphics[width=3.6in,height=2.5in]{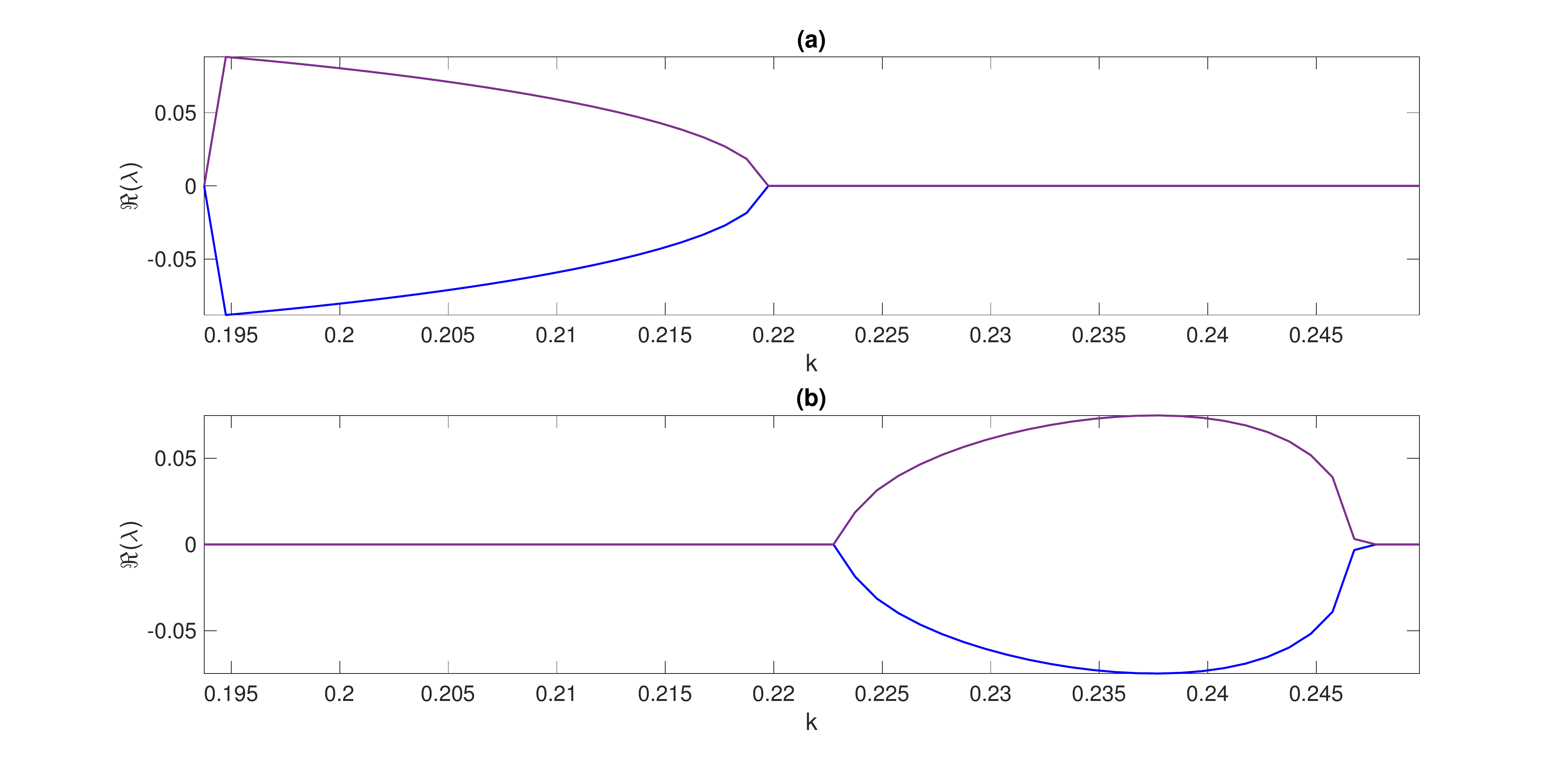}
        \caption{The real parts of the eigenvalues $(\lambda)$ corresponding to $P_1$ [Subplot (a)] and  $P_3$ [Subplot (b)]   are shown against the parameter $k$. The fixed parameter values are $\beta = 8$, $\alpha=0.15$, and  $n_0=10$. The   bifurcations indicate that the system [Eq. \eqref{eq-system}] can be  stable with $\Re\lambda<0$ or unstable with $\Re\lambda>0$  around the fixed point $P_1$ or $P_3$ in a finite domain of $k$.   }
  \label{fig:eigen}
\end{figure}

We numerically examine the roots of    Eq. \eqref{eq-char} within the domain $(2/\beta)\left(1-\alpha^2n_0\right)\lesssim k\lesssim (2/\beta)$ for some fixed values of   the other parameters, namely $\beta = 8$, $\alpha=0.15$, $n_0=10$.  Note that the qualitative features will remain the same for some other set   of   parameter values  fulfilling the restrictions for $\alpha,~n_0$ and $k$ stated before. Since we are interested in the real parts of the eigenvalues corresponding to the equilibrium points  $\left(x_{10},~x_{20},~0,~n\pi\right)$, without loss of generality, we assume that $n=0$.  
  The  real parts of the eigenvalues corresponding to   $P_1$ and $P_3$   are displayed in the subplots (a) and (b) of Fig.\ref{fig:eigen}. Note that the real eigenvalues corresponding to  $P_2$ and $P_4$ will remain the same as for  $P_1$ and $P_3$ respectively. Also, of four eigenvalues only two distinct  are shown for $P_1$ and $P_3$.  It is noted that depending on the ranges of values of $k$, the eigenvalues can assume zero, negative and positive values, indicating that the system can be stable (when $\Re\lambda$ is zero  or negative) or unstable (when $\Re\lambda>0$) about the equilibrium points. From the subplots (a) and (b), it is also seen that a critical value of $k$ exists near $k=0.22$, below or above which the system's stability may break down before it again reaches a steady state with a zero or a negative eigenvalue.  Since we have seen that the modulational instability of DAWs takes place in $0<k<k_c$, the   domain of $k$ in subplot (b) may provide an initial guess for the existence of chaos and periodicity in the wave-wave interactions.  
\subsection{Lyapunov exponents, bifurcation diagram and phase-space portraits}
Having predicted the stable and unstable regions of the dynamical system \eqref{eq-system} in the domains of the wave number of modulation $k$ as in Sec. \ref{sec-dynm-props}, we proceed to establish the ranges of values of the parameters $k$, $\alpha$, and $\beta$ in which the periodic, quasiperiodic or chaotic states of plasma waves can exist. To this end, we  first calculate the Lyapunov exponents $\Lambda_i$, $i=1,2,3,4$, for the  dynamical system [Eq. \eqref{eq-system}], to be written in the form $\dot{x}_i=f_i(X)$,  with the initial condition:  $X(0)=\left[x_1(0),~x_2(0),~x_3(0),~x_4(0)\right]$. We  are interested in the evolution of attractors and depending on the initial condition, these attractors will be associated with different sets of    exponents. The latter, however, describe the behaviors of $X(t)$ in the tangent space of the phase-space and are defined by the Jacobian matrix, given by,
 \begin{equation}
 J_{ij}(t)=\frac{df_i}{dx_j}\biggr|_{X(t)}.
\end{equation} 
The evolution of the tangent vectors can then be defined by the matrix $A$ via the following relation  
\begin{equation}
\dot{A}=JA,
\end{equation}
together with the initial condition $A_{ij}(0)=\delta_{ij}$. Here, $\delta_{ij}$ is the Kronecker delta and the matrix $A$ characterizes how a small change of separation distance between two trajectories in phase space develops from the starting point $X(0)$ to the final point $X(t)$. Nonetheless, the matrix $A$ is given by 
 \begin{widetext}
 \begin{equation}
 A(t) = \begin{bmatrix}
0 &\beta_0\sin x_4 &0 & px_2\cos x_4 \\
0 &0 &1 &0 \\
-\alpha^2_0\cos x_1\cos x_4& -k^2  & 0 &-\alpha_0^2\sin x_1\sin x_4\\
-\frac{1}{4}\beta_0 x_2\sec^2 (x_1/2)\cos x_4 &-\frac{1}{2}\beta_0\tan(x_1/2)\cos x_4& 0 &-\frac{1}{2}x_2\tan(x_1/2)\sin x_4
\end{bmatrix}.
 \end{equation}
 \end{widetext}
The Lyapunov exponents $\Lambda_i$ are thus obtained as  the eigenvalues of the following matrix.  
\begin{equation}
\Lambda = \lim_{t\rightarrow \infty}\frac{1}{2t}\log\left[A(t) A^T(t)\right], \label{eq-lyap}
\end{equation}
where $A^T$ denotes the transposed matrix of $A$. 
Given an initial condition $X(0)$, the separation distance between two trajectories in phase space or the change of particle's orbit can be obtained by the Liouville's formula: $\delta X(t)=\mathrm{tr}\left(J(t)\right)|A(t)|$, where $|A(t)|\equiv\mathrm{det}~A(t)$ and $\mathrm{det}~A(0)=1>0$. Thus, for the dynamical system \eqref{eq-system}, one obtains $\mathrm{det}~A(t)=\exp\left(\int_0^t\mathrm{tr}\left(J(t)\right)dt\right)=1>0$. It follows that at least one $\Lambda_i>0$, implying the existence of a chaotic state in a given time interval $[0,t]$.
\begin{figure}
\begin{center}
            \includegraphics[width=3.6in,height=2.5in]{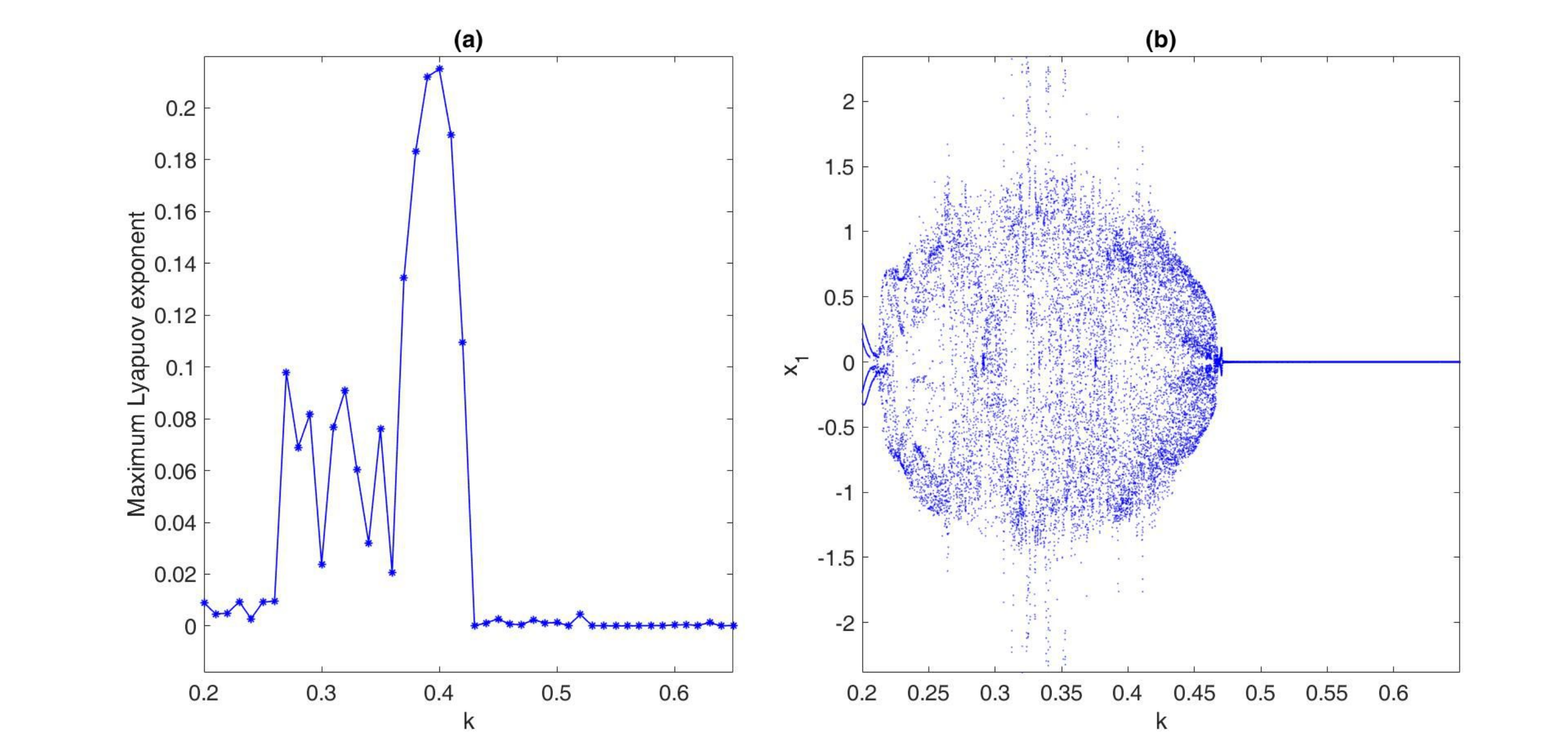}
        \caption{The maximum Lyapunov exponent [subplot (a)] and   the bifurcation diagram [subplot (b)] are shown against the wave number $k$. The  fixed parameter values are the same as in Fig. \ref{fig:eigen}. }
        \label{fig:lyap-bifur}
        \end{center}
\end{figure}
\par 
Before   proceeding further to the analyses of Lyapunov exponent spectra and the bifurcation diagram together with the phase-space portraits, we recapitulate that the MI of Alfv{\'e}n wave envelopes sets in for $0<k<k_c$. The growth rate of instability tends to become higher in the interval $0<k<(3/4)k_c$ and lower in   $(3/4)k_c<k<k_c$ with a cut-off at $k=k_c$ at which the pitchfork bifurcation occurs. It follows that the nonlinear dynamics of wave-wave interactions is subsonic in the interval $(3/4)k_c<k<k_c$. However, as $k$ decreases from  $(3/4)k_c$, many more unstable wave modes can be excited due to a selection of modes with $M=[k^{-1}]$ and the dynamics may no longer be subsonic. In this situation, a description of nonlinear interactions with three wave modes may be relatively correct. Thus, one may assume that one $(|m|=1)$   Alfv{\'e}n wave mode is unstable (i.e., the Alfv{\'e}n waves with $|m|>1$ are stable) and two driven ion-sound waves of plasma slow response (already excited by the unstable Alfv{\'e}n  mode)  remain as they are. This leads to the autonomous system \eqref{eq-system}. We will investigate how the system behaves as  the values of $k$ is successively increased from $(3/4)k_c$ to $k_c$ in the subsonic region and as $k$ reduces from $(3/4)k_c$  to a value so that the three-wave interaction model remains valid (since smaller the values of $k$ larger is the number of modes $M$). 
\par
\begin{figure}
            \includegraphics[width=3.6in,height=2.5in]{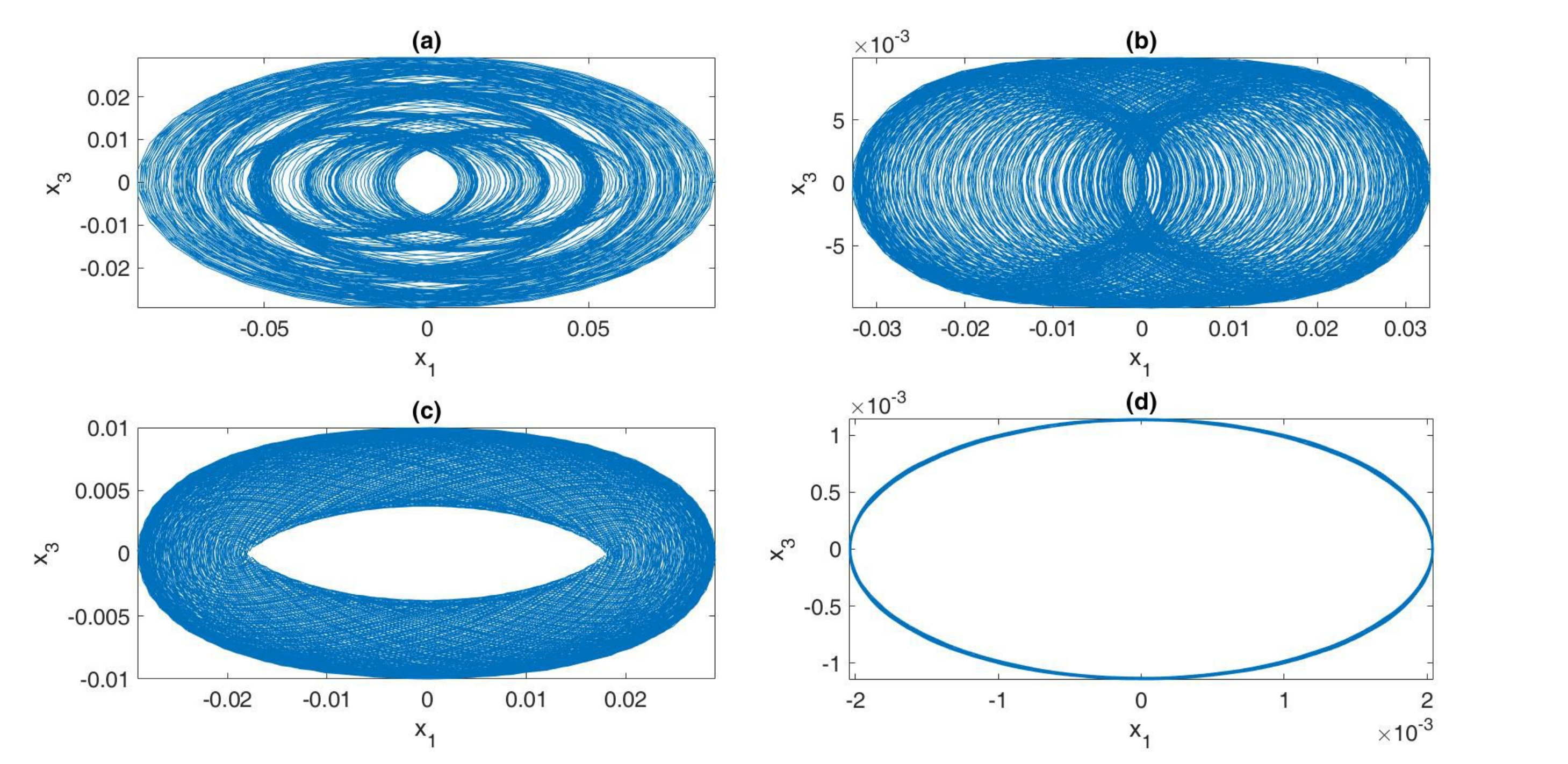}
\caption{Different phase-space portraits, formed in the dynamical evolution of Eq. \eqref{eq-system}, are shown for different values of the wave number of modulation $k$:   (a) $k=0.21$, (b) $k=0.42$, (c) $k=0.45$, and (d) $k=0.51$. Subplots (a) and (b) show the chaotic states while subplots (c) and (d) exhibit the quasi-periodic and periodic states respectively.  The fixed parameter values are the same as in Fig. \ref{fig:eigen}, i.e.,  $\beta = 8$, $\alpha=0.15$ and $n_0=10$. }
  \label{fig:phase-space}
\end{figure}

 
In what follows, we calculate the maximum Lyapunov exponent $\lambda_i^\text{max}$ for Eq. \eqref{eq-system} using the algorithm as stated above in a finite domain of $k$, i.e., $0<k<k_c<1$ and    numerically solve Eq. \eqref{eq-system}  using the fourth order Runge-Kutta scheme with a time step $dt=10^{-3}$ to obtain the bifurcation diagram of a state variable $x_1$ and phase-space portraits with the same set of fixed parameter values  $\beta = 8$, $\alpha=0.15$, and $n_0=10$ as in Fig. \ref{fig:eigen}. The results are displayed in Figs.  \ref{fig:lyap-bifur} and \ref{fig:phase-space}.  It is noted that, similar to subplot (b) of Fig. \ref{fig:eigen}, two sub-intervals of $k$ exist, namely  $0.2\lesssim k\lesssim k_1\approx0.42$ and $k_1<k\lesssim1$. In the former $\lambda_i^\text{max}>0$, while in the latter it is close to zero, implying that the system may exhibit chaotic states in $0.2\lesssim k\lesssim k_1$ and  quasiperiodic and/or limit cycles in the other sub-interval [See subplot (a) of Fig. \ref{fig:lyap-bifur}]. Physically, since lower (higher) values of $k~(<k_1)$ correspond to a large (small) number of wave modes ($[k^{-1}]$) to participate in the nonlinear wave dynamics, the wave-wave interactions may result into chaos (limit cycles or steady states) by the influence of the nonlinearity associated with the Alfv{\'e}n wave ponderomotive force (proportional to $\alpha$) and the nonlinear interactions between the fields (proportional to $\beta$).      These features can also be verified from the bifurcation diagram of a state variable, e.g., $x_1$ with respect to the parameter $k$ [See subplot (b) of Fig. \ref{fig:lyap-bifur}]. Here, it is seen that  as the value  of $k$ increases  within the domain, a transition from chaotic (dense region) to a periodic or steady (straight line) state can occur. However, the values of $k$ smaller than $k=0.2$ may not be admissible as those   correspond to a larger number of wave modes and their interactions   can not be described by the low-dimensional system \eqref{eq-system}  but by the full system of equations \eqref{eq3} and \eqref{eq4}.  An investigation of the latter is, however, out of scope of the present work. 
\par  
In order to further verify the dynamical features so predicted for ranges of values of $k$  and for illustration purpose, different phase-space portraits are also obtained by solving  Eq. \eqref{eq-system} numerically.  From Fig.   \ref{fig:phase-space} it is evident that as the values of $k$ increase  from   smaller [subplots (a) and (b)] to larger ones [subplots (c) and (d)], the   chaotic states of AWs transit into quasiperiodic [subplot (c)] and periodic [subplot (d)] states.  These features are in agreement with the Lyapunov exponent and the bifurcation diagram shown in Fig. \ref{fig:lyap-bifur}.  
\par
Thus, it is noted that the nonlinear interaction of a few wave modes of dispersive Alfv{\'e}n waves and low-frequency plasma density perturbations can exhibit    periodic, quasi-periodic and chaotic states in finite domains of the wave number of modulation due to the finite effects of the nonlinearities associated with the wave electric field driven ponderomotive  force and the interactions of the electric field and the plasma density fluctuations. The existence of these states is established by the analyses of Lyapunov exponents, the bifurcation diagram and the phase-space portraits. 
\section{Characterization of chaos: measure of chaos complexity}\label{sec-complexity}
 In this section, we  study the complexity of the dynamics of  wave-wave interactions and thereby measure quantitatively the characteristics of chaotic behaviors of the state variables relating to plasma density or wave electric field perturbations.  Although several formulas have been developed in the literature to characterize chaos, we focus mainly on the measures of embedding dimension estimation \cite{wallot2018}, correlation dimension \cite{she2019}, and the approximate entropy \cite{pincus1995,pincus1991,delgado2019}.
 \subsection{Estimation of Embedding parameters} 
 Many well known and efficient techniques, e.g., Recurrence quantification technique for the analysis of nonlinear time series require  the construction of phase-space profiles of the time series, since those techniques are applicable to the phase-space profiles  but not to the time series themselves. 
 The method of embedding dimension estimation is one such which also requires the reconstruction of successive phase spaces of   chaotic processes  with  the effects of time delay. 
\subsubsection*{Phase space reconstruction}
Reconstruction of phase space has become useful to extract information of a chaotic time series in nonlinear dynamical systems. Let $X=\left[x(1),~x(2),...,x(n)\right]^T$
  represent a uniformly sampled univariate time signal, i.e.,  an observed sequence of   the chaotic state variable $x(t)$ (which may be any one of $x_1,~x_2,~x_3$, and $x_4$) with $t=1,2,...,n$. Then to reconstruct a phase space by embedding the dimension $m$, we construct a time series $Y(t)$ of length $m$ (i.e., $m$-dimensional points) from the original time series $X(t)$ by considering   an appropriate time delay   $\tau$  as  
\begin{equation}
Y(t)=\left\lbrace X(t),~X(t+\tau),...,X[t+(m-1)\tau] \right\rbrace^T,
\end{equation}
where $t=1,2,...,n-(m-1)\tau$ and $\tau$ is a positive integer. Thus, the phase spaces  of $M=n-(m-1)\tau$ state variables are reconstructed. Generalizing this result,  one can reconstruct the phase spaces for  multivariate time signals. So, in order to perform the phase-space reconstruction, one must know   the  two embedding parameters, namely the time delay parameter $\tau$, which is the lag at which the time series has to be plotted against itself, and the embedding dimension parameter $m$, where $m-1$ is the number of times that the time series has to be plotted against itself using the delay $\tau$. Having known these two parameters,  one can  then reconstruct an approximate phase-space of the original one from a given time series. In the following two subsections, we estimate these two parameters  by the methods of computing the two functions, namely the Average mutual information (AMI) and the False nearest neighbors (FNN) \cite{wallot2018} in which the first local minima (or the points of cut-off) of these functions  can be estimated as the time delay and the embedding dimension respectively.

  \subsubsection*{Average mutual information (AMI): Estimation of time delay} 
  In AMI,  the mutual information is computed between the original    time series of a state variable   $X(t)$ and a time shifted version of the same time series, i.e., $X(t+\tau)$.   This average or  auto mutual information can be considered as a nonlinear generalization of the auto correlation function,  given by,
\begin{equation}
I\left[X(t),X(t+\tau)\right]=\sum_{i,j} p_{ij}(\tau)\log\left(\frac{p_{ij}(\tau)}{p_ip_j}\right),
\end{equation}
where $p_i$ is the probability that $X(t)$ is in the $i$th rectangle of the histogram  to be constructed  from the data points of $X(t)$ and $p_{ij}$ is the probability that $X(t)$ is in the $i$th rectangle and $X(t+\tau)$   in the $j$th rectangle.   
\par 
\subsubsection*{False nearest neighbors (FNN): Estimation of embedding dimension}
Typically, the embedding dimension $m$ for phase-space reconstruction is estimated by inspecting the change in distance between two nearest points in phase-space  as one gradually embeds the original time series $X(t)$ into higher dimensional ones $Y(t)$. The use of FNN, as prescribed by Kennel \textit{et al.} \cite{kennel1992}, is based on the following logic: Initially, we have the one-dimensional time series $X(t),~t=1,2,...,n$ and the distance between two of its neighboring points are noted.  Then   we embed   $X(t)$  into two dimensions $Y(t)=\left\lbrace X(t),~X(t+\tau)\right\rbrace$ with  some time delay  $\tau$ and examine whether there is any considerable change in the distance between any two neighboring data points of $Y(t)$. If so,  these data points are said to be  false neighbors,
and   the data points need  to be embedded further.  Otherwise,  if the change is not significant, the data points are called true neighbors and the embedding  retains the shape of the phase-space attractor, implying that   the present embedding dimension is sufficient. This process of successively increasing  the embedding dimension $m$  can be continued until the number of FNN reduces to zero, or the  subsequent embedding  does not alter the number of FNNs, or the   number of FNNs starts to increase again. A working algorithm for calculating FNN for our system can be stated below.
\begin{itemize}
\item[1] Identify the nearest point in the Euclidean sense  to a
given point of the time-delay coordinates. That is, for a given
time series $Y(t)=\left\lbrace X(t),~X(t+\tau),...,X[t+(m-1)\tau]\right\rbrace^T$,
find a point $Y_i$ in the data set such that the  
distance   $m=\|Y_i-Y_j\|_2$ is minimized, where $Y_i$ and $Y_j$ denote the nearest neighboring data points of $Y(t)$.
\item[2] Determine whether the following expression is true or false:
 \begin{equation}
 \frac{|X_{i}-X_{j}|}{\|Y_i-Y_j\|}\leq \mathrm{Distance ~threshold~(R)},\label{threshold}
 \end{equation}
 where $X_i$ and $X_j$ denote the nearest neighboring data points of $X(t)$.
 If the condition in Eq. \eqref{threshold} is satisfied, then the neighbors are true nearest neighbors, otherwise they are false nearest neighbors.
 \item[3] Perform the step 2   for all  points $i$ in the data set and calculate the percentage of points in the data set that have false nearest neighbors.
\item[4] Increase embedding the dimension   until the percentage of false nearest neighbors drops to zero or an admissible small number.
\end{itemize}
\par
\begin{figure} 
\begin{center}
            \includegraphics[width=3.5in,height=2.5in]{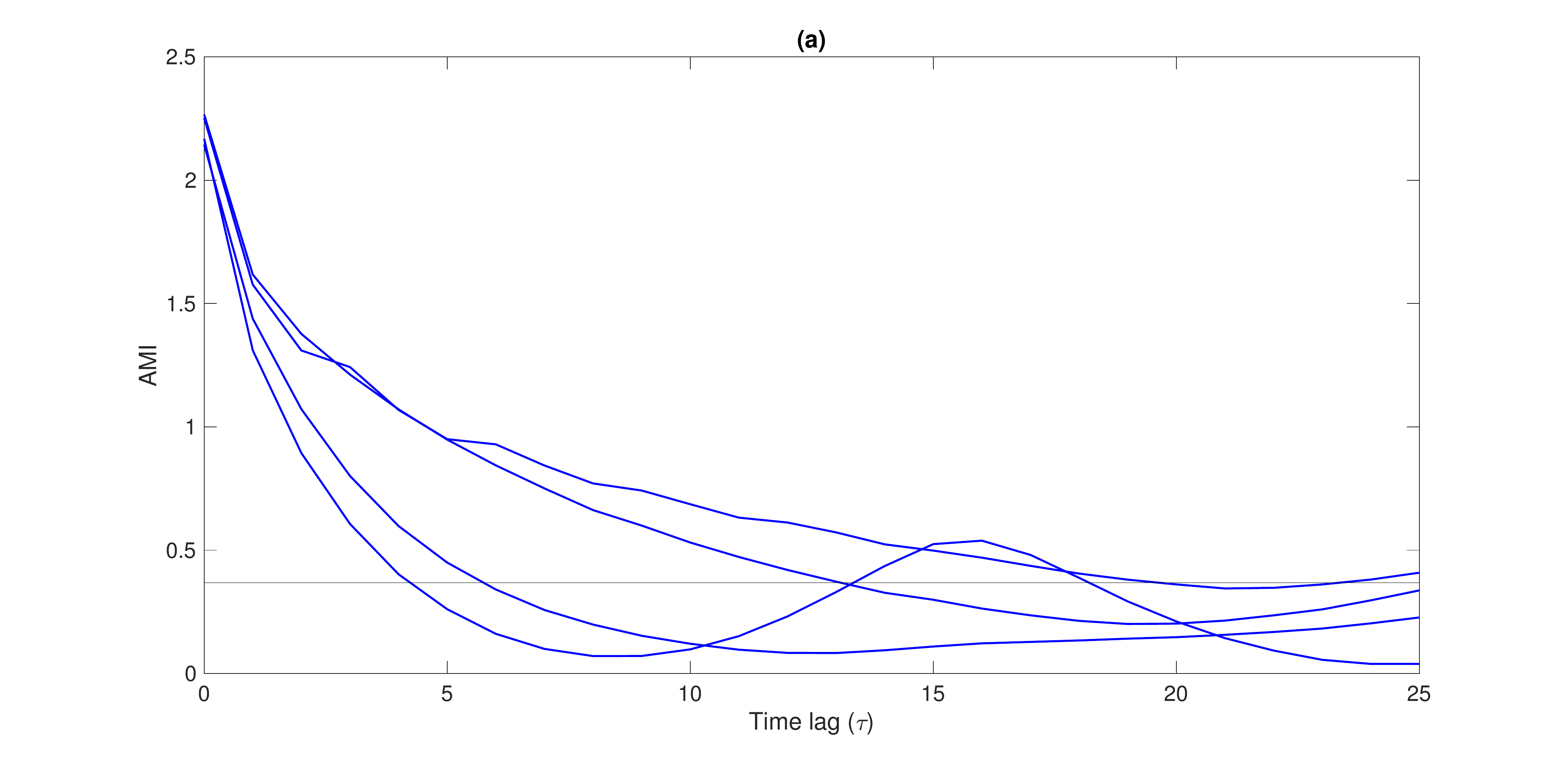}
            \includegraphics[width=3.5in,height=2.5in]{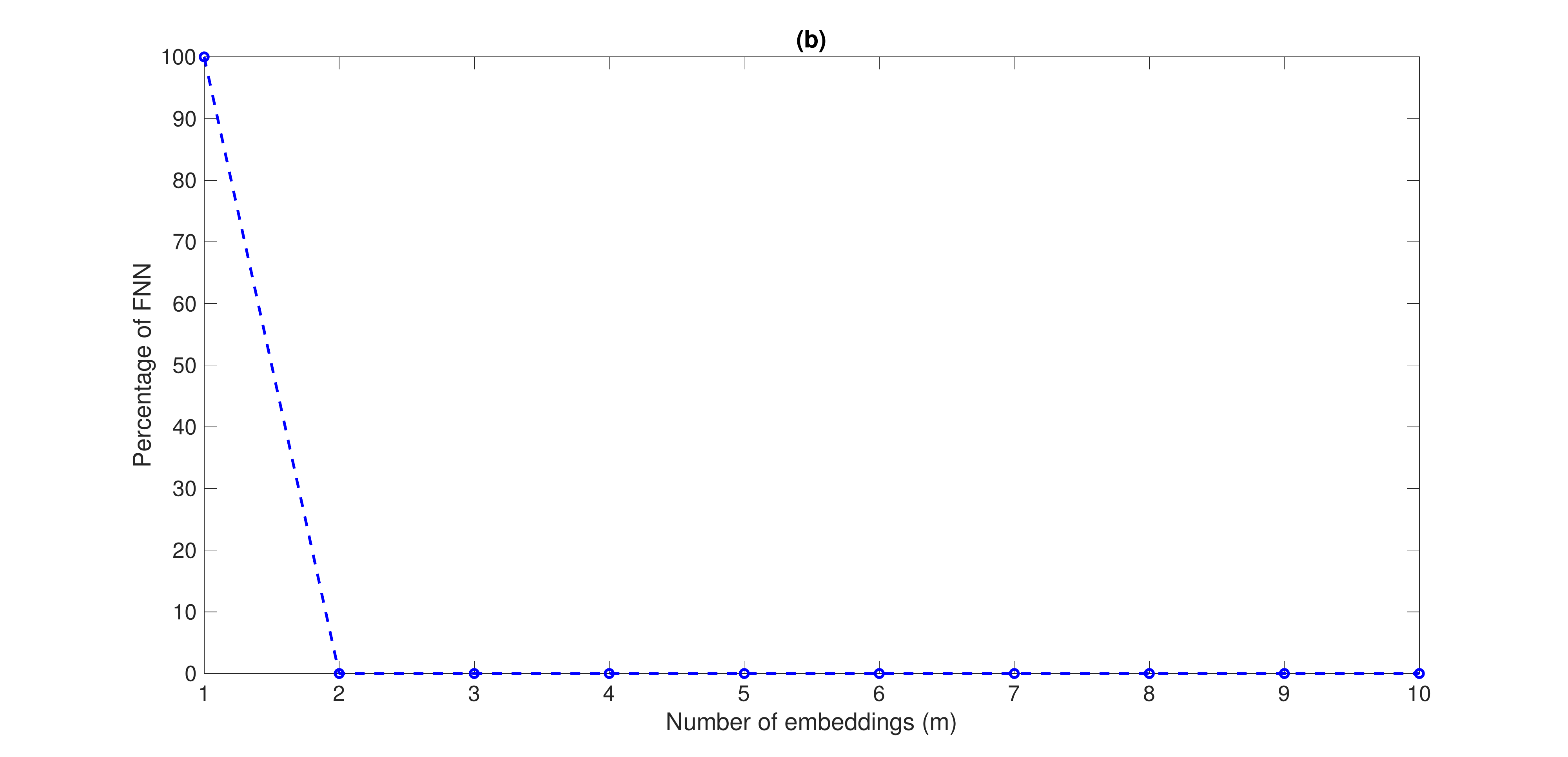}
            
        \caption{    The graphical output(s)   of the average mutual information (AMI) function [Subplot (a)] and percentage of false nearest neighborhoods (FNN) function [Subplot (b)]  are shown against the time lag $\tau$ and the embedding dimension $m$ respectively for the four-dimensional time series taken from  Eq. \eqref{eq-system}. In the upper panel, the default threshold value $(1/e)$ is shown by the horizontalline.   The dashed line in the lower panel shows an immediate drop-off of the
percentage of false-nearest neighbors to zero, indicating that no additional embedding is necessary for the  time series.   }
        \label{fig:embp}
        \end{center}
\end{figure}

\begin{figure*} 
\begin{center}
            \includegraphics[width=4.5in,height=3in]{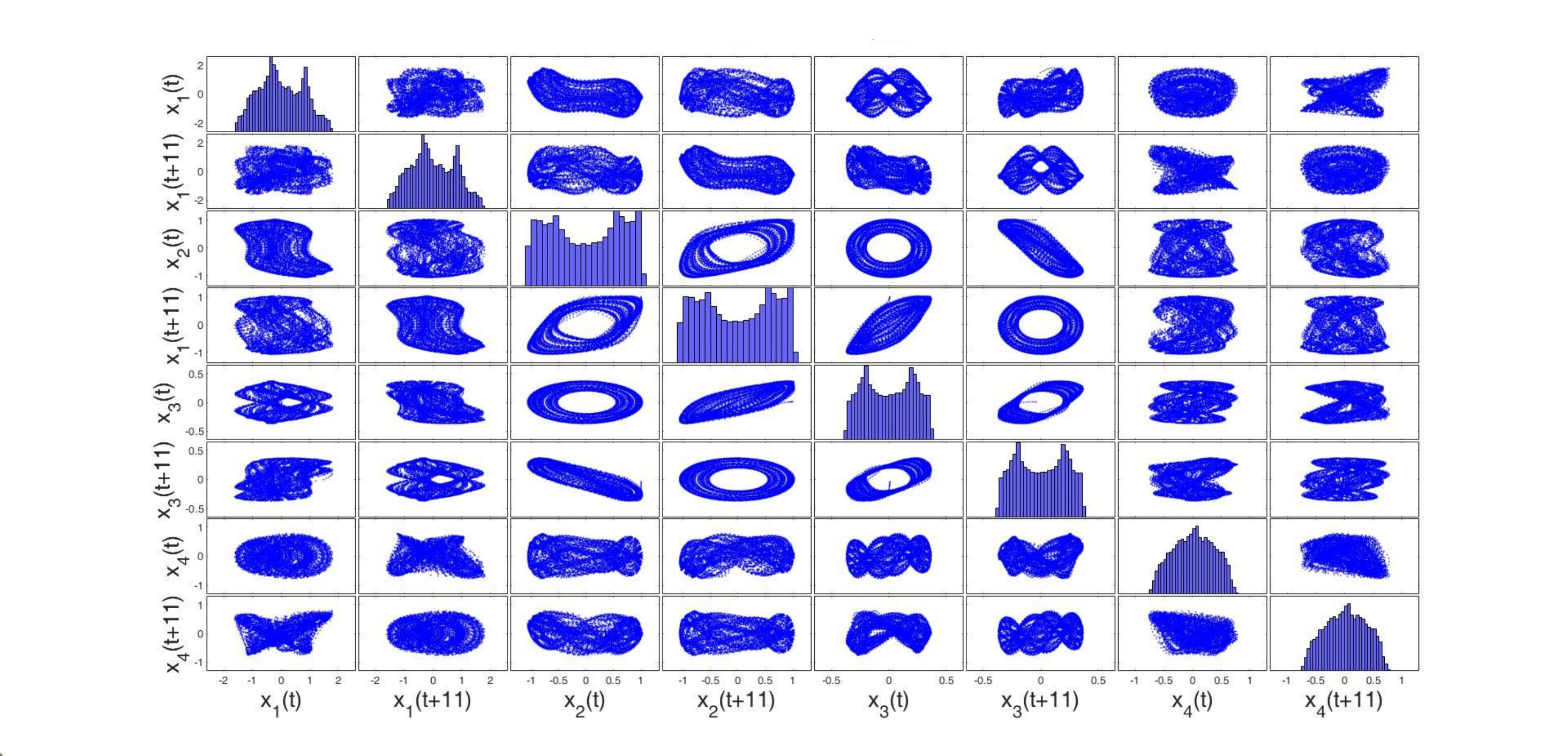}
        \caption{ Reconstruction of phase-space with the time lag, $\tau=11$ and embedding dimension, $m=4$ from the chaotic time series of Eq. \eqref{eq-system}. }
        \label{fig:reconst}
        \end{center}
\end{figure*}

 
Following Ref. \cite{wallot2018} and using MATLAB, we estimate the embedding parameters, namely the time lag $\tau$ and the embedding dimension $m$ for the four-dimensional time series $(x_1,~x_2,~x_3,~x_4)$ formed by all the four variables of the system \eqref{eq-system}.  
The results are shown in Fig. \ref{fig:embp}. From subplot (a), we find that all the  auto mutual information (AMI) curves, obtained for different time series,   cut the threshold line at different values of $\tau$.  It is seen that for these curves, the AMI first drops below the threshold value $(1/e)$ after the time lags $\tau=4.25,~5.75,~13.1$, and $19.66$, and we have considered the maximum time delay as $\tau=25$. Thus, a mean value of $\tau$ for each dimension can be obtained as $\tau=(4.25+5.75+13.1+19.66)/4=10.7$ for which we obtain an estimate for $\tau$ as $\tau=11$.  On the other hand, subplot (b) displays the  FNN function against the embedding dimension $m$. It is clear that the available four-dimensional time series is sufficient and no further time-delayed embedding of dimension is required.   
Next, having estimated the time delay $\tau=11$ and the embedding dimension $m>2$ a reconstruction of phase space is shown in Fig. \ref{fig:reconst} for all the variables $x_1$, $x_2$, $x_3$, and $x_4$ of Eq. \eqref{eq-system}. Here, the time series for the variables are plotted against each other with the time lag $\tau=11$. It is seen that the resulting phase space with the time lag is approximately the same as the original one. 
\subsection{Correlation dimension estimation}
One of the most important measures of complexity of chaotic attractors is the correlation (or fractal) dimension. It has been shown by many researchers that the correlation  dimension is more pertinent to experimental data than the capacity dimension as it simply calibrates the geometrical structure of an attractor and is insufficient for higher dimensional systems. Moreover, the correlation dimension is a (or close to the) lower bound on the Hausdorff fractal dimension, which is infinite for noise;  positive and finite for a deterministic system; integer for integrable systems, and non-integer for a chaotic deterministic system. The derivation of the correlation dimension also requires the reconstruction of vectors from the time series $X(t)$, i.e., 
\begin{equation}
\begin{split}
Y_1(t)&=\left\lbrace X(t),~X(t+\tau),...,X[t+(m-1)\tau] \right\rbrace^T,\\
Y_2(t)&=\left\lbrace X(t+p),...,X[t+p+(m-1)\tau] \right\rbrace^T,\\
&\vdots\\
Y_M(t)&=\left\lbrace X(t+Mp),...,X[t+Mp+(m-1)\tau] \right\rbrace^T,
\end{split}
\end{equation}
where  $p$ (a positive integer) and $\tau$, respectively, stand  for the inter-vector and intra-vector spacing.
\par
After the reconstruction of phase space of a chaotic signal $X(t)$ with $M$ vectors and computing the correlated  vector pairs, its proportion in all possible pairs in $M^2$ is the correlation integral $C(l)$, given by,
\begin{equation}
C(l)=\lim_{M\rightarrow\infty}\left[\frac{2}{M^2}\sum_{i=1}^{M-k}\sum_{j=i+k}^{M}\Theta\left(l-|X(i)-X(j)| \right) \right], \label{eq-corri}
\end{equation}
where $X(i)$ and $X(j)$ are the position vectors of pints on an attractor, $l$ is the distance under  inspection, $k$ is the summation offset  used to prevent proximate vectors   being counted, and $\Theta(x)$ is the Heaviside step function, defined by,
\begin{equation}
\Theta(x)=\left\lbrace 
\begin{array}{cc}
0,& x\leq0\\
1,& x>0.
\end{array}\right.
\end{equation}
The correlation dimension $d$ is then calculated from the correlation integral as
\begin{equation}
d=\lim_{l\rightarrow0} \frac{\log {C(l)}}{\log{l}}. \label{eq-corrd}
\end{equation}  
\par
\begin{figure*}
\begin{center}
            \includegraphics[width=4.5in,height=2.5in]{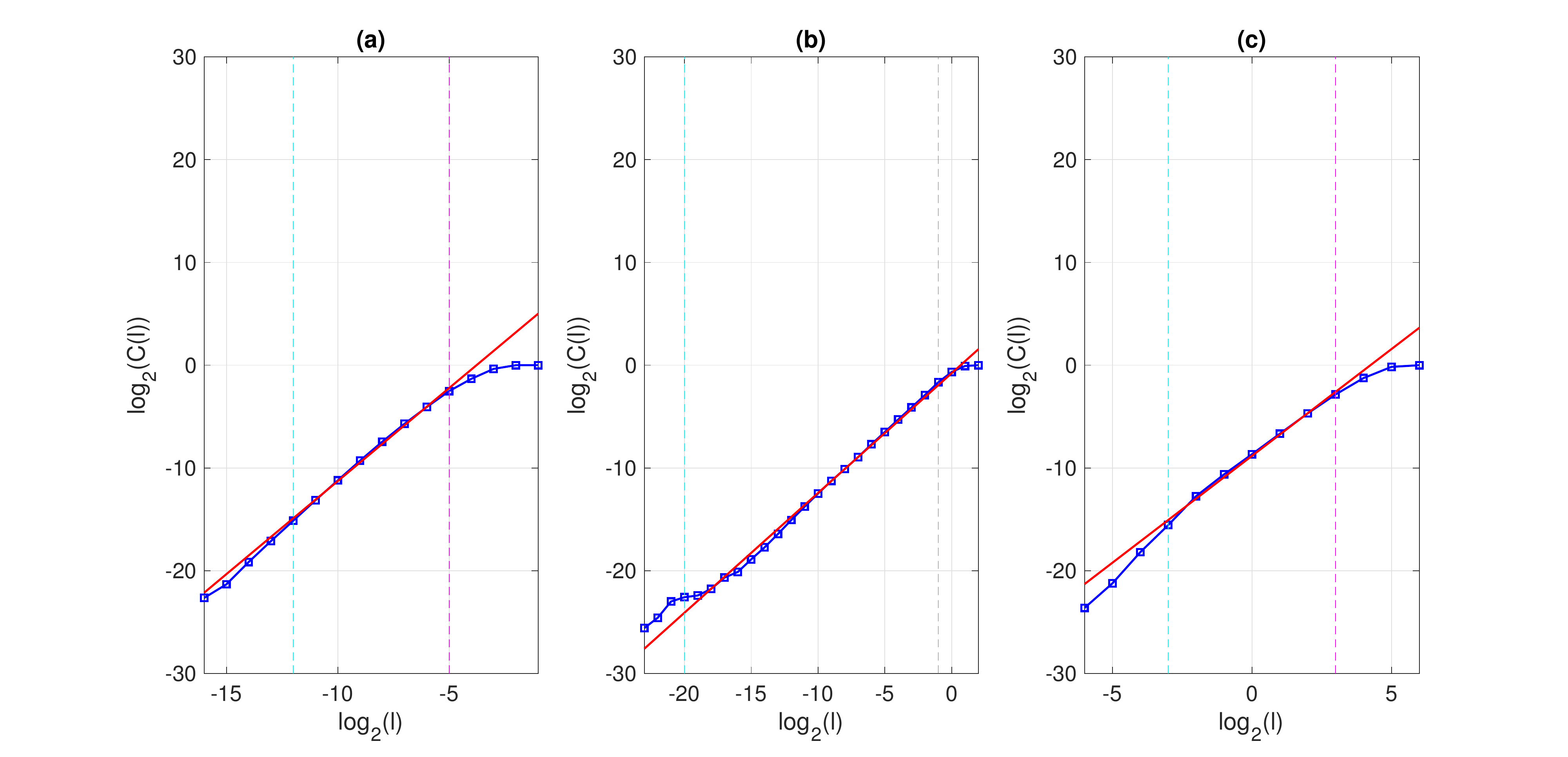}
        \caption{  Plots of the log-correlation integral $[\log{C(l)}]$ vs the log-correlation dimension   $(\log{l})$  are shown for (a)  the present model [Eq. \ref{eq-system}], (b)   for the H{\'e}non map, (c) for the Lorenz system.   The correlation dimensions obtained for  the subplots  (a), (b), and (c)  are   $d=1.0776$,  $1.25$, and  
$2.06$ respectively. }
        \label{fig:corr1}
        \end{center}
\end{figure*}

 
Next, using  Eq. \eqref{eq-corrd} we plot a graph  of $\log{C(l)}$ vs $\log{l}$ for the time series $X(t)$ of the dynamical system \eqref{eq-system} with  a fixed embedding dimension $m=4$ and time delay $\tau=11$. The system is turned to be higher dimensional by the method described above.  As a comparison we have also obtained graphs of the correlation integrals for the H{\'e}non map with the embedding dimension $m=2$ and the Lorenz system with   $m=3$.  The results are displayed in Fig. \ref{fig:corr1}.  The fixed parameter values   considered here are the same as   Fig. \ref{fig:eigen}, i.e., $\alpha=0.15$, $\beta=8$, and $n_0=10$. The slopes of the straight-line  portions (obtained using the least-square curve fitting) of the graphs represent the correlation dimensions. For the present system [Eq. \eqref{eq-system}] the results as in subplot (a) appear similar to those for the H{\'e}non map [subplot (b)] and the Lorenz system [subplot (c)]. However, the correlation dimension obtained for our system is $d=1.0776$, while for the H{\'e}non map and the Lorenz system are $d=1.25$ and $d=2.06$ respectively. It follows that the system \eqref{eq-system} is chaotic and possesses a strange attractor characterized by  $d=1.0776$.  
\par
\begin{figure*}
\begin{center}
            \includegraphics[width=4.5in,height=2.5in]{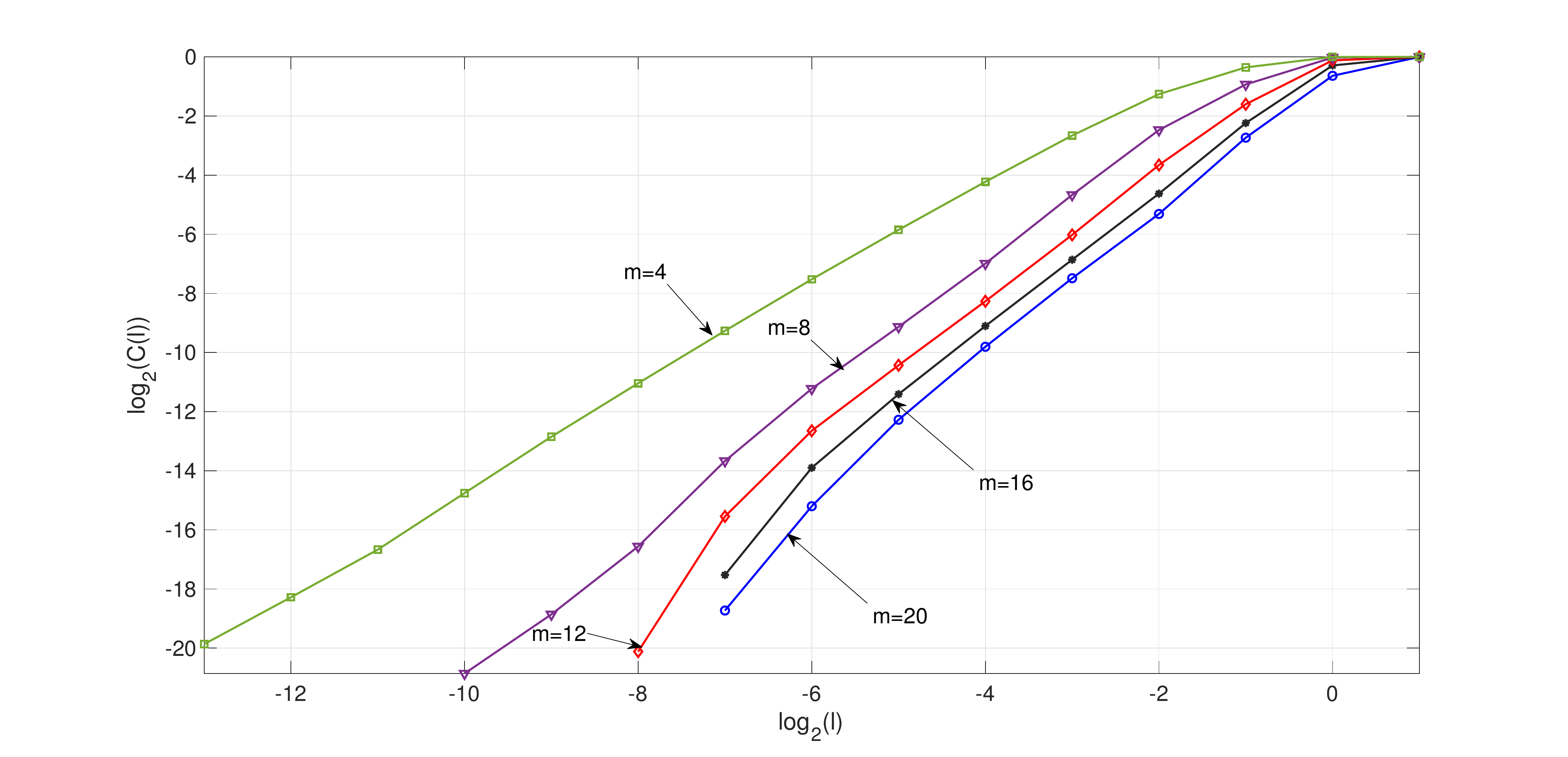}
        \caption{Plot of the log-correlation integral $[\log{C(l)}]$ vs the log-correlation dimension   $(\log{l})$  with different embedding dimensions: $m=4,~8,~12,~16,~20$ and $10,000$ observations  is shown for  Eq.\eqref{eq-system}. The parameter values are the same as Fig. \ref{fig:eigen}, i.e.,  $\beta = 8$, $\alpha=0.15$, and $n_0=10$ together with $k=0.35$. }
        \label{fig:corr2}
        \end{center}
\end{figure*}

 
In Fig. \ref{fig:corr2}, we plot $\log{C(l)}$ vs $\log{l}$    for increasing values of the embedding dimension, namely   $m=8,~12,~16$, and $20$. The time series is taken to be consisted of $10000$ points separated by the time lag $\tau=11$.  One can then obtain the correlation dimensions as  $d=1.0777,~1.0781,~1.0788$ respectively.   Thus,   a series of straight lines indeed exist with  slopes $d\approx1.07\pm0.01$ and  $\log{C(l)/\log{l}}$ is nearly a constant value for large $m$. 
\begin{widetext}
\begin{center}
\begin{table}[h]
\begin{tabular}{|p{0.5in}|p{0.4in}|p{0.4in}|p{0.4in}|p{0.4in}|p{0.4in}|p{0.4in}|p{0.6in}|} \hline 
  Model & Control parameter & \multicolumn{5}{|p{1.9in}|}{Correlation dimension(d) for different values of $k$ and $m$ } &  Hausdorff fractal dimension (D) \\ \cline{3-7}
 &    & d \newline(m=4) & d \newline (m=8) & d \newline (m=12) & d \newline (m=16) & d \newline (m=20) &  \\ \hline
Our Dynamical Model  &  k=0.13 & 1.0776 & 1.0777 & 1.0781 & 1.0788 & 1.0811 &  $1.0646 \pm 0.013997$ \\ 
 &  k=0.23 & 1.0453 & 1.0465 & 1.0473 & 1.0483 & 1.0535 &  \\ 
 &  k=0.28 & 1.0718 & 1.0724 & 1.0733 & 1.0740 & 1.0747 &  \\  
 &  k=0.33 & 1.0720 & 1.0705 & 1.0694 & 1.0686 &  1.0767 &  \\ 
 &  k=0.38 &   1.0721 & 1.0706 &  1.0696 & 1.0684 & 1.0677 &  \\\cline{2-8}
 &  k=0.5 & 1.00069 & 1.00068 & 1.00065 & 1.00063 & 1.00061 & $1.00047 \pm 0.00068262$ \\ \hline 
Henon map &  a=1.4 & 1.2541 & 1.4850 & 1.6445 & 2.0881 & 3.1845 &  $1.260303 \pm 0.003$ \\  
 &  a=1.3 & 1.2510   & 1.4516 & 1.6565 & 2.1834 & 2.6294 &  \\ 
 &  a=1.25 & 1.2506 & 1.4615 & 1.6756 & 2.1225 & 2.6252 &  \\ 
 &  a=1.2 & 1.2501 & 1.4589 & 1.6946 & 2.1045 & 2.6052 &  \\  
 &  a=1.15 & 1.2408 & 1.4568 & 1.6725 & 2.0768 & 2.6466 &  \\ 
 &  a=1.1 & 1.2659 & 1.4588 & 1.6767 & 2.04794 & 2.6246 &  \\ \hline
 Lorenz system &  $\rho=28$ & 2.0866 & 2.7763 & 3.0121 & 3.1246 & 3.3623 & 2.06 $\mathrm{\pm}$ 0.01 \\  
 &  $\rho=26$ & 2.1670 & 2.3767 & 2.5679 & 2.6844 & 2.8563 &  \\  
 &  $\rho=24$ & 1.8288 & 2.0223 & 1.9516 & 2.1235 & 2.4874 &  \\  
 &$\rho=22$ & 2.1999 & 2.0413 & 1.9608 & 2.0915 & 2.4599 &  \\\cline{2-8}
 &  $\rho=20$ & 0.01047 & 0.0779 & 0.0478 & 0.0148 & 0.0048 & $2.0002 \pm 0.00048023$ \\ 
 &  $\rho=15$ & 0.0079 & 0.0076 & 0.0075 & 0.0072 & 0.0071 &  \\ \hline 
\end{tabular}
\caption{Estimations for the correlation dimension $(d)$ and the Hausdroff fractal dimension $(D)$ are shown  for the present model, the H{\'e}non map and the Lorenz system  with different values of $m$ and the control parameter $k$ and with a fixed $10000$ observations.}
\label{table1}
\end{table}
\end{center}
\end{widetext}
\par In what follows, we   calculate the correlation dimension $(d)$ and the Hausdroff fractal dimension $(D)$ [For details, see, e.g., Ref. \cite{mori1980}] of a time series of Eq. \eqref{eq-system} with   different values of the control parameter $k$ and the embedding dimension $m$. The results are compared with those of the H{\'e}non map and the Lorenz system. A summary of the results is presented in  Table \ref{table1}.   It is noted that even with an increasing value of the embedding dimension and a change of value of the parameter $k$, the correlation dimension $d$ converges to a constant value. The bounds for the Hausdroff dimension of the chaotic time series are also calculated. It is seen that the correlation dimension lies within the bounds of the Hausdroff dimension.    

\subsection{Approximate entropy (ApEn)}
Although a number of techniques are used to measure the complexity of a chaotic system, not all are applicable to limited, noisy and stochastically derived time series. For example, the Kolmogorov-Sinai (KS) entropy works well for real dynamical systems but not for systems with noise \cite{delgado2019}. Also, the finite correlation dimension value discussed before can not guarantee that the system under consideration is deterministic. Furthermore, the Pincus technique fails for systems dealing with stochatic components.  In this situation, the  Approximate Entropy (ApEn) is more applicable to measure the system's complexity  compared to others in which  the statistical precision   is compromised \cite{delgado2019}. The ApEn estimates uniformly sampled time-domain signals through phase space reconstruction and then measures the amount of regularity and unpredictability of fluctuations in a time series. 
%
\begin{figure} 
\begin{center}
            \includegraphics[width=3.6in,height=2.5in]{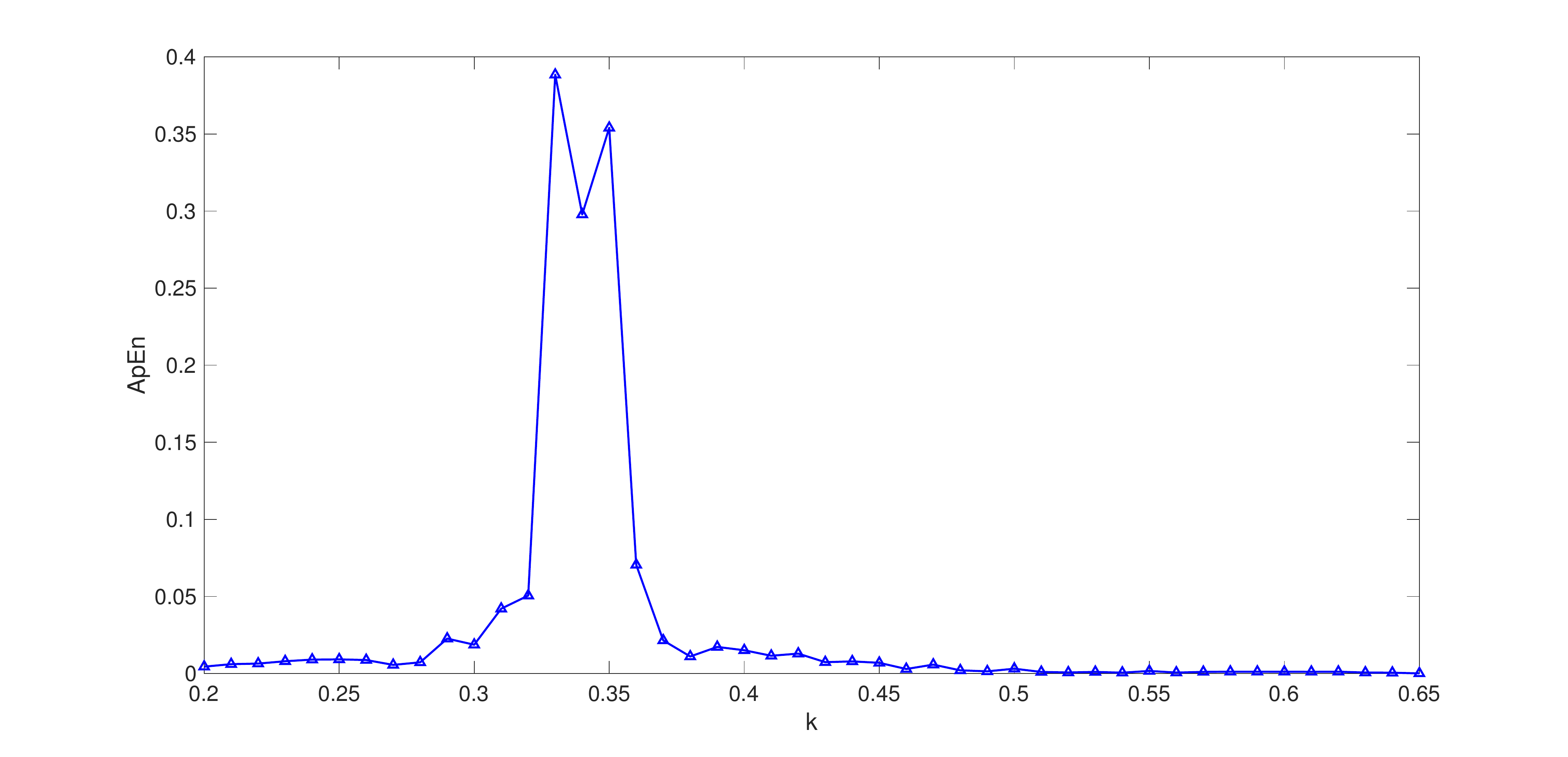}
        \caption{Approximate entropy (ApEn) is shown against the parameter $k$  for Eq. \ref{eq-system} with $m=4$,   $l=0.1$,  and $N=5000$. The other parameter values are the same as Fig. \ref{fig:eigen}, i.e., $\beta = 8$, $\alpha=0.15$, and $n_0=10$. }
        \label{fig:ApEn}
        \end{center}
\end{figure}
 For  an $N$ given data points together with the embedding dimension $m$  and the correlation integral $C(l)$, the ApEn is defined by
\begin{equation}
{\rm{ApEn}}(m,l,N)=\Phi^m(l)-\Phi^{m+1}(l),
\end{equation}
where
\begin{equation}
\Phi^m(l) = \frac{\sum_{i=1}^{N-m+1}\log C^m_i(l)}{N-m+1}. 
\end{equation}
We have calculated the ApEn against the controlling parameter $k$ (The other parameter values are the same as Fig. \ref{fig:eigen}, i.e.,  $\beta = 8$, $\alpha=0.15$, and $n_0=10$) and for a  given set of values, namely  $m=4$ and $l=0.1$ together with   $5000$ data points. The results are shown graphically in Fig. \ref{fig:ApEn}. It is noted that while the ApEn assumes high values in the subdomain $0.275\lesssim k\lesssim0.4$ in which the Lyapunov exponent is found to be positive (\textit{cf}. Fig. \ref{fig:lyap-bifur}) its values are low in rest of the domain where the Lyapunov exponents are close to zero. Thus, low values of ApEn   predict that  the system is steady, tedious and predictive,  while high values imply the independence between the data, a low number of repeated patterns, and randomness.   
   
\section{Conclusion} \label{sec-conclu}
We have investigated the dynamical properties of dispersive Alv{\'e}n waves coupled to plasma slow response of electron and ion density perturbations in a uniform magnetoplasma. By restricting the nonlinear wave-wave interactions to a few  numbers of active wave modes, a low-dimensional autonomous system is constructed which is shown to exhibit periodic, quasiperiodic and chaotic states by means of the analyses of Lyapunov exponent spectra,   bifurcation diagram, and   phase-space portraits.  The low-dimensional autonomous system can be a good approximation for the nonlinear interaction of  Alfv{\'e}n waves coupled to driven ion-sound waves associated with plasma slow response of density fluctuations in the stable or plane wave region $(3/4)k_c<k<k_c<1$. In the latter,  the modulational instability growth rate of Alv{\'e}n wave envelopes is low.  The model can be relatively accurate in the region $0.2\lesssim k<(3/4)k_c$ (in which the condition for the subsonic region is relaxed and the instability growth rate is relatively high) where the low-dimensional model exhibits chaos for  given values of the the pump electric field $E_0$ as well as the parameters $\alpha$, $\beta$, and $n_0$, associated with the relative speeds of the Alv{\'e}n waves compared to the speed of light in vacuum and the ion-sound speed, and the conserved plasmon number respectively. However, for values of $k<0.2$, the low-dimensional model will no longer be valid for the description of wave-wave interactions as smaller values of $k$ correspond to the excitation of a large number of unstable modes.
\par 
The complexity of chaotic phase-space structures of chaotic time series is also measured quantitatively by means of the correlation dimension and the approximate entropy through the reconstruction of phase spaces and estimation of embedding parameters, namely the time lag   and the embedding dimension. It is found that even with an increasing value of the embedding dimension and with a slightly different set of values of   the parameters $\alpha$, $\beta$,   $n_0$, and $k$, the correlation dimension converges to a constant value. The bounds for the Hausdroff fractal dimension of the chaotic time series are also calculated to show that the correlation dimension lies in between the bounds. Furthermore, the results are shown to be a good qualitative agreement with those   for the H{\'e}non map and the Lorenz system.
\par
To conclude, the existence of chaos and its complexity in the low-dimensional interaction model can be a good signature for the emergence of spatiotemporal chaos in the full system of equations \eqref{eq1} and \eqref{eq2} where the  participation of many more wave modes (more than three)  in the nonlinear interactions can be possible. Such chaotic aspects   of Alv{\'e}n waves can be relevant for the onset of turbulence due to flow of energy from lower to higher harmonic modes (i.e., with large to small spatial length scales) in the Earth's ionosphere and magnetosphere.  
\begin{acknowledgments}
 The authors thank the anonymous reviewers for their valuable comments. A. Roy and A. P. Misra wish to thank SERB (Government of India) for support through a research project with sanction order no. CRG/2018/004475.  
\end{acknowledgments}
\section*{Data Availability} The data that support the findings of this study are available from the corresponding author upon reasonable request.
\appendix 

\section{Derivation of the dispersion relation [Eq. \eqref{eq-disp}]} \label{appendix-A}
Here, we give some relevant details for the derivation of the dispersion relation \eqref{eq-disp} for the modulated DAW envelope. 
We rewrite Eqs. \eqref{eq3} and \eqref{eq4} as
 \begin{equation}
\frac{\partial^2 n}{\partial t^2}-\frac{\partial^2 n}{\partial z^2}= -\alpha^2 \frac{\partial^2}{\partial z^2} |E|^2,											\label{eq-a1}
\end{equation}
\begin{equation}
\frac{\partial E}{\partial t}+\beta \frac{\partial E}{\partial z}-\frac{\beta
}{2} \frac{\partial}{\partial z}(nE)+i \gamma \frac{\partial ^2 E}{\partial z^2}=0,	\label{eq-a2}
\end{equation}
We assume the wave electric field envelope to be of the form $E=\tilde{E}e^{i\theta(z,t)}$ and the density perturbation as $n=\tilde{n}(z,t)$. Then  Eqs. \eqref{eq-a1} and \eqref{eq-a2} reduce to  
\begin{equation}
\frac{\partial^2 \tilde{n}}{\partial t^2}-\frac{\partial^2 \tilde{n}}{\partial z^2}= -\alpha^2 \frac{\partial^2}{\partial z^2} |\tilde{E}|^2,											\label{eq-a3}
\end{equation}
\begin{equation}
\begin{split}
\frac{\partial \tilde{E}}{\partial t}+&i\tilde{E} \frac{\partial \theta}{\partial t}+ \beta\left(\frac{\partial \tilde{E}}{\partial z}+ i\tilde{E} \frac{\partial \theta}{\partial t}\right)-\frac{\beta}{2}\left[\frac{\partial}{\partial z}(\tilde{n}\tilde{E})+i\tilde{E} \tilde{n} \frac{\partial \theta}{\partial z}\right]\\
& +i \gamma \left[\frac{\partial^2 \tilde{E}}{\partial z^2}+ 2i \frac{\partial \tilde{E}}{\partial z} \frac{\partial \theta}{\partial z}+ i \tilde{E} \frac{\partial^2 \theta}{\partial z^2}- \tilde{E}\left(\frac{\partial \theta}{\partial z}\right)^2 \right]=0,						 \label{eq-a4}
\end{split}
\end{equation}
Separating the real and imaginary parts of Eq. \eqref{eq-a4}, we get 
\begin{equation}
\frac{\partial \tilde{E}}{\partial t}+ \beta \frac{\partial \tilde{E}}{\partial z}- \frac{\beta}{2} \frac{\partial}{\partial z}(\tilde{n} \tilde{E})- \gamma \left(\tilde{E} \frac{\partial^2 \theta}{\partial z^2}+ 2 \frac{\partial \tilde{E}}{\partial z} \frac{\partial \theta}{\partial z}\right)=0,												   \label{eq-a5}
\end{equation}  
\begin{equation}
\tilde{E} \frac{\partial \theta}{\partial t}+ \beta \tilde{E} \frac{\partial \theta}{\partial z}- \frac{\beta}{2} \tilde{E} \tilde{n} \frac{\partial \theta}{\partial z}+ \gamma \left[\frac{\partial^2 \tilde{E}}{\partial z^2}- \tilde{E}\left(\frac{\partial \theta}{\partial z}\right)^2\right]=0.\label{eq-a6}
\end{equation}
Looking for the modulation of the Alfv{\'e}n wave envelope, we make the following ansatz: 
\begin{equation} \label{eq-ansatz1}
\begin{split}
\tilde{E}(z,t)= E_0+E_1 \cos{(kz-\omega t)}+E_2 \sin{(kz-\omega t)},\\
\tilde{n}(z,t)=n_1 \cos{(kz-\omega t)}+n_2 \sin{(kz-\omega t)},\\
\theta(z,t)=\theta_0+\theta_1 \cos{(kz-\omega t)}+ \theta_2 \sin{(kz-\omega t)}, 
\end{split}
\end{equation}
where $E_0,~ E_1,~ E_2,~ n_1,~ n_2,~ \theta_0, ~\theta_1$, and $ \theta_2$ are real constants. 
\par 
Substituting   Eq. \eqref{eq-ansatz1} into Eqs.  \eqref{eq-a3}, \eqref{eq-a5}, and \eqref{eq-a6} and linearizing (retaining only the first harmonic terms), we get 
\begin{equation}
\begin{split}
&\left[(\omega^2-k^2)n_1+ 2 \alpha^2 k^2 E_0 E_1\right]\cos{(kz-\omega t)}\\
&+ \left[(\omega^2-k^2)n_2+2 \alpha^2 k^2 E_0 E_2\right] \sin{(kz-\omega t)}=0,         \label{eq-a8}
\end{split}
\end{equation}
\begin{equation}
\begin{split}
&[(\omega-\beta k)E_2+ \frac{\beta}{2} E_0 k n_2-\gamma E_0 k^2 \theta_1]\cos{(kz-\omega t)}\\
&-[(\omega-\beta k)E_1+ \frac{\beta}{2} E_0 k n_1+\gamma E_0 k^2 \theta_2]\sin{(kz-\omega t)}=0,               														 \label{eq-a9}
\end{split}
\end{equation}
\begin{equation}
\begin{split}
&[(\omega-\beta k)E_0 \theta_2+ \gamma k^2 E_1]\cos{(kz-\omega t)}\\
&-[(\omega-\beta k)E_0 \theta_1-\gamma k^2 E_2]\sin{(kz-\omega t)}=0.   \label{eq-a10}
\end{split}
\end{equation}
Equating the coefficients of different harmonics proportional to $\cos{(kz-\omega t)} $ and $ \sin{(kz-\omega t)} $ to zero,   we successively obtain
\begin{equation}
(\omega^2-k^2)n_1+2\alpha^2k^2E_0E_1=0,				 \label{eq-a11a}
\end{equation}
 \begin{equation}
(\omega^2-k^2)n_2+2\alpha^2k^2E_0E_2=0, 				\label{eq-a11b}
\end{equation}
\begin{equation}
(\omega-\beta k)E_2+\frac{\beta}{2}E_0 k n_2-\gamma E_0 k^2 \theta_1=0,	 \label{eq-a12a}
\end{equation}
\begin{equation}
(\omega-\beta k)E_1+\frac{\beta}{2}E_0 k n_1+\gamma E_0 k^2 \theta_2=0,  \label{eq-a12b}
\end{equation}
\begin{equation}
(\omega-\beta k)E_0\theta_2+\gamma k^2 E_1=0,			 \label{eq-a13a}
\end{equation}
\begin{equation}
(\omega-\beta k)E_0\theta_1-\gamma k^2 E_2=0.			 \label{eq-a13b}
\end{equation}
Next, eliminating $\theta_1$ and $\theta_2$ from Eqs. \eqref{eq-a12a}-\eqref{eq-a13b}, we get
\begin{equation}
[(\omega-\beta k)^2-\gamma^2 k^4] E_1+ \frac{\beta}{2} (\omega-\beta k) E_0 k n_1=0 ,				 																			\label{eq-a14a}
\end{equation}  
\begin{equation}
[(\omega-\beta k)^2-\gamma^2 k^4] E_2+ \frac{\beta}{2} (\omega-\beta k) E_0 k n_2=0. 				 																					\label{eq-a14b}
\end{equation}
Furthermore, eliminating either $n_1$ from Eqs. \eqref{eq-a11a} and \eqref{eq-a14a} or eliminating $n_2$ from  Eqs. \eqref{eq-a11b} and \eqref{eq-a14b}, and noting that $E_1,~ E_2\neq0$, we obtain
\begin{equation}
(\omega^2-k^2)[(\omega-\beta k)^2-\gamma^2 k^4]-\alpha^2 \beta k^3 |E_0|^2 (\omega-\beta k)= 0.																		\label{eq-a15}
\end{equation}
From Eq. \eqref{eq-a15}, it is noted that while the first term represents a coupling between the Alfv{\'e}n wave and the ion-acoustic density perturbation, the second term proportional to $|E_0|^2$ appears due to the Alfv{\'e}n wave driven ponderomotive force. In absence of the latter, we have the usual acoustic mode $\omega=k$ and the following Alfv{\'e}n wave dispersion equation.
\begin{equation}
\omega-\beta k=-\gamma k^2, 											\label{eq-a16}
\end{equation}
where the negative sign (on the right-hand side) is considered in order to satisfy Eq. \eqref{eq-a2} for the wave eigenmode. 
So, treating the term proportional to $|E_0|^2$ as the correction term in Eq. \eqref{eq-a15} and replacing $(\omega-\beta k )$ by $-\gamma k^2$ therein, we obtain the following dispersion law for the modulated Alfv{\'e}n wave envelope.
\begin{equation}
(\omega^2-k^2)[(\omega-\beta k)^2-\gamma^2 k^4]+ \alpha^2 \beta \gamma k^5 |E_0|^2=0.
																					  \label{eq-a17}
\end{equation}
Next, to obtain the growth rate of instability, we assume $\omega \approx \beta k+i\Gamma$ with $\beta k \gg\Gamma,~ \gamma k^2$, Thus,   Eq. \eqref{eq-a17} gives
\begin{equation}
[(\beta^2-1)k^2-\Gamma^2+2i\beta k \Gamma](\Gamma^2+\gamma^2 k^4)-\alpha^2\beta \gamma k^5 |E_0|^2=0.												 \label{eq-a18}
\end{equation}
Since the   term proportional to $i$ in Eq. \eqref{eq-a18} does not give any admissible result, we   equate the real part to zero. Thus, we obtain  
\begin{equation}
[(\beta^2-1)k^2-\Gamma^2](\Gamma^2+\gamma^2 k^4)-\alpha^2 \beta \gamma k^5 |E_0|^2=0.																			 \label{eq-a19}
\end{equation}
Using $\beta k\gg\gamma k^2$ and neglecting the terms containing higher orders (than the second order) of $\Gamma$,  we obtain from Eq. \eqref{eq-a19} the following expression for the growth rate of instability.  
\begin{equation}
\Gamma^2=\gamma k^3\left(\frac{\beta \alpha^2 |E_0|^2}{\beta^2-1}-\gamma k\right). \label{eq-a20}
\end{equation}
\section{Derivation of the low-dimensional model [Eqs. \eqref{e7}-\eqref{e9}]  } \label{appendix-B}
We recast Eqs. \eqref{eq3} and \eqref{eq4} as
\begin{equation}
\frac{\partial^2 n}{\partial t^2}-\frac{\partial^2 n}{\partial z^2}= -\alpha^2 \frac{\partial^2}{\partial z^2} |E|^2,											\label{eq-b1}
\end{equation}
\begin{equation}
\frac{\partial E}{\partial t}+\beta \frac{\partial E}{\partial z}-\frac{\beta
}{2} \frac{\partial}{\partial z}(nE)+i \gamma \frac{\partial ^2 E}{\partial z^2}=0.	\label{eq-b2}
\end{equation}
Next, we consider an one dimensional spectrum for each of the the wave electric field $E$ and the plasma density perturbation $n$, which describe the general solution of Eqs. \eqref{eq-b1} and \eqref{eq-b2}  as a  superposition  of a set of normal modes, i.e., 
\begin{equation}
E(z,t)=E_0(t)+E_{-1}(t). e^{-ik z}+E_{1}(t). e^{ik z},		\label{eq-b3}
\end{equation} 
\begin{equation}
n(z,t)=n_0(t)+n_{1}(t). e^{ik z}+n_{1}^*(t). e^{-ik z},		\label{eq-b4}
\end{equation}
where $E_{-1}(0)=E_{1}(0)$. 
\par
Substituting  these expressions for $E$ and $n$ into \eqref{eq-b2} and following the same approach as in Refs. \cite{banerjee2010,misra2010}, we obtain 
\begin{equation}
i\dot{E}_0=0,					\label{eq-b5}			
\end{equation}
\begin{equation}
\dot{E}_1+ik \beta E_1-ik ^2 \gamma E_1=i \frac{k \beta}{2}(n_0E_1+n_1E_0), \label{eq-b6}	
\end{equation}
\begin{equation}
\dot{E}_{-1}-ik \beta E_{-1}-ik ^2 \gamma E_{-1}=-i \frac{k \beta}{2}(n_0E_{-1}+n_1^*E_0), \label{eq-b7} 
\end{equation}
where the dot   denotes the differentiation with respect to $t$ and the asterisk denotes the complex conjugate.
Multiplying Eq \eqref{eq-b5} by $E_0^*$ we obtain
\begin{equation}
i|\dot{E_0}|^2=0.																\label{eq-b10}
\end{equation} 
Also, multiplying Eqs. \eqref{eq-b6} and \eqref{eq-b6} successively by $E_1^*$ and $E_{-1}^*$ and subtracting the complex conjugate of the resulting equations from themselves, we get  
\begin{equation}
i|\dot{E_{-1}}|^2=\frac{k \beta}{2}(n_1^*E_{0}E_{-1}^*-n_1E_{0}E_{-1}),		\label{eq-b8} 
\end{equation}
\begin{equation}
i|\dot{E_{1}}|^2=\frac{k \beta}{2}(n_1^*E_{0}E_{1}-n_1E_{0}E_{1}^*),		\label{eq-b9}
\end{equation}
where  $\dot{|E|}^2=\frac{d}{dt}|E|^2$. Equations \eqref{eq-b10}-\eqref{eq-b9} can be added to yield
\begin{equation}
|E_{-1}|^2+|E_{1}|^2+|E_0|^2=N.													\label{eq-b11}
\end{equation}
Next, we assume   $n_1=n_1^*,~ n_0=N$, the plasmon number, and introduce  the new variables $\rho_{0},~ \rho_{1},~ \theta_{0}$, and $\theta_{1}$ according to 
 $E_0=\rho_{0}e^{i\theta_{0}},E_{-1}=E_0=\rho_{1}e^{i\theta_{1}},\rho_0=\sqrt{n_0}\sin{w},\rho_1=\sqrt{n_0}\cos{w}, \psi=2w$.
  \par 
Substituting the expressions \eqref{eq-b3} and \eqref{eq-b4} into Eq. \eqref{eq-b1} and using the new  variables as defined above, we get 
\begin{equation}
\ddot{n_1}-k^2 n_1=\alpha^2 k^2 n_0\sin{\psi}\cos{\phi}.						\label{eq-b12}
\end{equation} 
Also, using the newly defined varibles, from Eqs. \eqref{eq-b6} and \eqref{eq-b7}  we obtain									\label{eq-b13}
\begin{equation}
\dot{\psi}=\beta k n_1 \sin{\phi}	,												\label{eq-b14}
\end{equation}
\begin{equation}
\dot{\phi}= k(\beta -  \gamma k)-\frac{1}{2} \beta k n_1 \tan{\frac{\psi}{2}} \cos\phi. \label{eq-b16}
\end{equation} where $\phi=\theta_0-\theta_1$. Equations \eqref{eq-b12}-\eqref{eq-b16} constitute the required low-dimensional model.
 \section{Equilibrium points of Eq. \eqref{eq-system}   } \label{appendix-C}
 To find the equilibrium points of Eq. \eqref{eq-system}, we equate the right-hand side expression of each  of Eq. \eqref{eq-system} to zero. Thus, we successively obtain  
\begin{equation}
\begin{split}
& \beta_0 x_2 \sin{x_4}=0,	\\
& x_3=0,\\
& -k^2 x_2+\alpha_0^2\sin{x_1}\cos{x_4}=0,	\\					 			 
&  \gamma_0-\frac{1}{2} \beta_0 x_2 \tan\frac{x_1}{2} \cos{x_4}=0,   \label{eq-c1}
\end{split}
\end{equation}
where $\beta_0=\beta k$, $\alpha_0=\alpha k \sqrt{n_0}$ and $\gamma_0=k(\beta -\gamma k)$.  
Since $(0,0,0,0)$ is not an equilibrium point (as explained in Sec. \ref{sec-eigen}),  we have $x_2 \neq 0$. 
So, the first equation of Eq. \eqref{eq-c1} gives $\sin{x_4}=0$, i.e., $x_4=n\pi$. Using this value of $x_4$ in the third equation of Eq. \eqref{eq-c1}, one obtains $x_2=(-1)^n(\alpha_0^2/k^2)\sin{x_1}$, since $\cos(n\pi)=(-1)^n$, $n$ being zero or an integer. Having obtained the  values of $x_2$ and $x_4$, and using those in the fourth equation of Eq. \eqref{eq-c1},  we get  
$x_{1}=4n\pi \pm2\sin^{-1}\left(\sqrt{{\gamma_0k^2}/{\alpha_0^2 \beta_0}}\right)$.
 Thus, the equilibrium points of Eq. \eqref{eq-system} can be obtained as   $(x_{10},x_{20},0,n\pi)$, where
$x_{10}=4n\pi \pm2\sin^{-1}\left(\sqrt{{\gamma_0k^2}/{\alpha_0^2 \beta_0}}\right)$ and $x_{20}=\pm(-1)^n\left({2k}/{\beta_0}\right)\sqrt{\left(\beta-\gamma k\right)\left(\alpha^2n_0\beta-\beta+\gamma k\right)}$, where $n$is zero or    an integer.  
 \bibliographystyle{apsrev4-1}
\bibliography{References}

\end{document}